\documentclass[12pt]{article} % default is 10pt
\usepackage[utf8x]{inputenc} % to set input encoding (not needed with XeLaTeX)
\usepackage[T1]{fontenc} % to set font encoding in latex; difference noticed for <>
\usepackage{amsmath,amssymb,bbm,commath} % for mathematical symbols
\usepackage{slashed} % for Dirac slash
\usepackage[unicode]{hyperref} % for bookmarks in Acrobat Reader and using hyperlinks
\hypersetup{bookmarksnumbered=true, bookmarksopen=true, bookmarksopenlevel=2, breaklinks=true, citecolor=blue, colorlinks=true, linkcolor=red, linktoc=page, pdfborder={0 0 0}, pdfkeywords={Semichirals, }{N=(2,2) Multiplets, }{Localization}, pdfstartview=FitH, pdfauthor={FBenini, PMCrichigno, DJain}, pdfsubject={Semichiral Fields on Sphere}, pdftitle={Semichiral Fields on S² and Generalized Kähler Geometry}, plainpages=false, unicode=true, urlcolor=cyan}
\usepackage{cite} % for compact citations
\usepackage{geometry} % to change the page dimensions
\usepackage{fancyhdr} % This should be set AFTER setting up the page geometry
\usepackage{parskip} % needed to not screw up 'other' spacings!
\usepackage[nottoc,notlof,notlot]{tocbibind} % to put the bibliography in the ToC
\usepackage[titles]{tocloft} % to alter the style of the ToC

\DeclareUnicodeCharacter{"0393}{\Gamma}
\DeclareUnicodeCharacter{"0394}{\Delta}
\DeclareUnicodeCharacter{"0398}{\Theta}
\DeclareUnicodeCharacter{"039B}{\Lambda}
\DeclareUnicodeCharacter{"039E}{\Xi}
\DeclareUnicodeCharacter{"03A0}{\Pi}
\DeclareUnicodeCharacter{"03A3}{\Sigma}
\DeclareUnicodeCharacter{"03A5}{\Upsilon}
\DeclareUnicodeCharacter{"03A6}{\Phi}
\DeclareUnicodeCharacter{"03A8}{\Psi}
\DeclareUnicodeCharacter{"03A9}{\Omega}
\DeclareUnicodeCharacter{"03B1}{\alpha}
\DeclareUnicodeCharacter{"03B2}{\beta}
\DeclareUnicodeCharacter{"03B3}{\gamma}
\DeclareUnicodeCharacter{"03B4}{\delta}
\DeclareUnicodeCharacter{"03B5}{\epsilon}
\DeclareUnicodeCharacter{"03B6}{\zeta}
\DeclareUnicodeCharacter{"03B7}{\eta}
\DeclareUnicodeCharacter{"03B8}{\theta}
\DeclareUnicodeCharacter{"03D1}{\vartheta}
\DeclareUnicodeCharacter{"03B9}{\iota}
\DeclareUnicodeCharacter{"03BA}{\kappa}
\DeclareUnicodeCharacter{"03BB}{\lambda}
\DeclareUnicodeCharacter{"03BC}{\mu}
\DeclareUnicodeCharacter{"03BD}{\nu}
\DeclareUnicodeCharacter{"03BE}{\xi}
\DeclareUnicodeCharacter{"03C0}{\pi}
\DeclareUnicodeCharacter{"03C1}{\rho}
\DeclareUnicodeCharacter{"03C3}{\sigma} %\sigma
\DeclareUnicodeCharacter{"03C4}{\tau}
\DeclareUnicodeCharacter{"03C5}{\upsilon}
\DeclareUnicodeCharacter{"03C6}{\phi}
\DeclareUnicodeCharacter{"03D5}{\varphi}
\DeclareUnicodeCharacter{"03C7}{\chi}
\DeclareUnicodeCharacter{"03C8}{\psi}
\DeclareUnicodeCharacter{"03C9}{\omega}
\DeclareUnicodeCharacter{"21D0}{\Leftarrow}
\DeclareUnicodeCharacter{"0212B}{\AA}
\DeclareUnicodeCharacter{"00B7}{\cdot}
\DeclareUnicodeCharacter{"266A}{\eighthnote}
\DeclareUnicodeCharacter{"266B}{\twonotes}
\PrerenderUnicode{²}\PrerenderUnicode{ä}

\newcommand{\Title}[1] {\title{\Huge #1}}
\newcommand{\TPheader}[3] {\thispagestyle{fancy}\pagenumbering{alph}\lhead{#1}\chead{#2}\rhead{#3}\cfoot{}}
\newcommand{\makepage}[1] {\newpage\pagenumbering{#1}}%\cfoot{\thepage}\rhead{}\chead{}\lhead{}}

\newcommand{\Abstract}[1] {\begin{abstract}\small #1 \end{abstract}}

\newcommand{\bea}{\begin{equation}\begin{aligned}}
\newcommand{\eea}{\end{aligned}\end{equation}}

\newcommand\eqs[1] {\begin{align}#1\end{align}}
\newcommand\eqsn[1] {\begin{align*}#1\end{align*}}

\newcommand\eqsg[1] {\eqs{\begin{split}#1\end{split}}}
\newcommand\equ[1] {\begin{equation}#1\end{equation}}
\newcommand\equn[1] {\begin{equation*}#1\end{equation*}}

\newcommand\pmat[1] {\begin{pmatrix}#1\end{pmatrix}}
\newcommand\spmat[1] {\big( \begin{smallmatrix} #1 \end{smallmatrix} \big)}
\newcommand\tabl[2] {\begin{table}[#1]\centering #2\end{table}}
\renewcommand\( {\left(}
\renewcommand\) {\right)}
\newcommand\wb {\overline}

\newcommand\wt {\widetilde}
\newcommand\e {\epsilon}
\newcommand\bep {{\bar{\epsilon}}}
\newcommand\s {\sigma}
\newcommand\unit {\mathbbm{1}}
\DeclareMathOperator{\Tr}{Tr}
\DeclareMathOperator{\Det}{Det}
\DeclareMathOperator{\coIm}{coIm}
\renewcommand\L {{\cal L}}
\newcommand\M {{\cal M}}
\newcommand\N {{\cal N}}
\renewcommand\O {{\cal O}}
\renewcommand\P {{\cal P}}
\newcommand\Q {{\cal Q}}
\renewcommand\S {{\cal S}}
\newcommand\X {{\cal X}}

\renewcommand\Bbb {\mathbb}
\newcommand\bD {{\Bbb D}}
\newcommand\bQ {{\Bbb Q}}
\newcommand\bR {{\Bbb R}}
\newcommand\bX {{\Bbb X}}
\newcommand\bZ {{\Bbb Z}}
\newcommand\fm {\mathfrak{m}}
\newcommand\fR {\mathfrak{R}}
\newcommand\sT {{\sf T}}
\newcommand\ie {\textit{i.e.}}

\newcommand\etc {\textit{etc.}}

\numberwithin{equation}{section} % Equations numbered as <section>.#
\begin{document}

\Title{Semichiral Fields on $S^2$ \\ and Generalized K\"{a}hler Geometry}

\author{Francesco Benini$^\diamondsuit$\footnote{\href{mailto:f.benini@imperial.ac.uk}{f.benini@imperial.ac.uk}}\,, P. Marcos Crichigno$^\spadesuit$\footnote{\href{mailto:p.m.crichigno@uu.nl}{p.m.crichigno@uu.nl}}\,, Dharmesh Jain$^{\natural}$\footnote{\href{mailto:djain@phys.ntu.edu.tw}{djain@phys.ntu.edu.tw}}\,, and Jun Nian$^{\clubsuit}$\footnote{\href{mailto:jnian@insti.physics.sunysb.edu}{jnian@insti.physics.sunysb.edu}}
\medskip\\
\small \emph{$\scriptstyle{^{\diamondsuit}}$Simons Center for Geometry and Physics} \\[-.4em]
\small \emph{State University of NewYork, Stony Brook, NY 11794, USA} \\
\small \emph{KITP, University of California, Santa Barbara, CA 93106, USA} \\
\small \emph{Delta Institute for Theoretical Physics, University of Amsterdam} \\[-.4em]
\small \emph{Science Park 904, 1098 XH Amsterdam, the Netherlands} \\
\small \emph{Blackett Laboratory, Imperial College London} \\[-.4em]
\small \emph{South Kensington Campus, London SW7 2AZ, United Kingdom}
\medskip\\
\small \emph{$\scriptstyle{^{\spadesuit}}$Institute for Theoretical Physics and} \\[-.4em]
\small \emph{Center for Extreme Matter and Emergent Phenomena} \\[-.4em]
\small \emph{ Utrecht University, Utrecht 3854 CE, the Netherlands}
\medskip\\
\small \emph{$^\natural$Center for Theoretical Sciences, Department of Physics} \\[-.4em]
\small \emph{National Taiwan University, Taipei 10617, Taiwan}
\medskip\\
\small \emph{$\scriptstyle{{}^{\spadesuit}}{{}^\natural}\scriptstyle{^{\clubsuit}}$C. N. Yang Institute for Theoretical Physics} \\[-.4em]
\small \emph{State University of New York, Stony Brook, NY 11794, USA}
}

\date{} % Displays a given date or no date (if empty), otherwise the current date is printed
\maketitle

\TPheader{\today}{}{YITP-SB-15-09\\ITP-UU-15/07\\Imperial/TP/2015/FB/02\\NSF-KITP-15-050}

\Abstract{We study a class of two-dimensional $\N=(2,2)$ supersymmetric gauge theories, given by semichiral multiplets coupled to the usual vector multiplet. In the UV, these theories are traditional gauge theories deformed by a gauged Wess-Zumino term. In the IR, they give rise to nonlinear sigma models on noncompact generalized K\"ahler manifolds, which contain a three-form field $H$ and whose metric is not K\"ahler. We place these theories on $S^2$ and compute their partition function exactly with localization techniques. We find that the contribution of instantons to the partition function that we define is insensitive to the deformation, and discuss our results from the point of view of the generalized K\"ahler target space.
}

\makepage{Roman} % Comment this line to make ToC appear on the title page.
\tableofcontents
\makepage{arabic}

\section{Introduction}\label{Introduction}

Starting with the work of Pestun \cite{Pestun:2007rz}, there has been substantial application of localization techniques \cite{Witten:1988xj, Witten:1991zz} to supersymmetric field theories in various dimensions, leading to the exact computation of Euclidean path-integrals on various manifolds and backgrounds. This new wave of exact results has affected two-dimensional physics \cite{Benini:2012ui, Doroud:2012xw, Gomis:2012wy, Gadde:2013dda, Benini:2013nda, Sugishita:2013jca, Honda:2013uca, Hori:2013ika, Benini:2013xpa, Doroud:2013pka, Kim:2013ola, Murthy:2013mya, Ashok:2013pya, Benini:2014mia, Gomis:2014eya, Benini:2015noa, Closset:2015rna}, in particular in the study of two-dimensional $\N=(2,2)$ gauge theories \cite{Benini:2012ui, Doroud:2012xw}.  One of the most interesting applications in two dimensions is to gauged linear sigma models (GLSMs) with chiral multiplets, realizing nonlinear sigma models (NLSMs) on K\"{a}hler manifolds in the infrared (IR) \cite{Witten:1993yc}. In this case, the exact computation of field theory observables leads to interesting quantities associated to the corresponding K\"{a}hler target spaces. For instance, in (untwisted) A-type localization the $S^2$ partition function computes the exact K\"{a}hler potential on the quantum K\"{a}hler moduli space of Calabi-Yau manifolds \cite{Jockers:2012dk, Gomis:2012wy}, while in B-type localization it computes the exact K\"{a}hler potential on the complex structure moduli space \cite{Doroud:2013pka}. Among other applications, this has been used to compute Gromov-Witten invariants \cite{Jockers:2012dk, Honma:2013hma} and the Seiberg-Witten prepotential in novel ways \cite{Park:2012nn}. The extension of localization techniques to theories involving twisted chiral multiplets has shed new light on mirror symmetry, showing that the $S^2$ partition function for the Landau-Ginzburg models proposed by Hori and Vafa \cite{Hori:2000kt, Hori:2003ic} reproduces the partition function of the corresponding mirror Abelian GLSMs \cite{Gomis:2012wy}.

In this paper we extend the localization techniques to more general $\N=(2,2)$ gauge theories by including \emph{semichiral multiplets}. A salient feature of these GLSMs is that they include a gauged Wess-Zumino term and they realize NLSMs on generalized K\"ahler manifolds in the IR, rather than K\"ahler when only chiral (or only twisted chiral) multiplets are present.  In fact, chiral, twisted chiral, and semichiral multiplets are all required (and sufficient) to describe the most general $\N=(2,2)$ NLSMs with torsion. The gauge theories we consider here are the $S^2$ versions of the GLSMs discussed in detail in \cite{Crichigno:2015pma}.\footnote{For some previous work on GLSMs with semichiral multiplets in flat space see \cite{Merrell:2006py, Lindstrom:2007vc, Lindstrom:2007sq, Merrell:2007sr, Crichigno:2011aa, CrichignoThesis}; for a study of gauge theories with a gauged Wess-Zumino term with on-shell $\N=(2,2)$ supersymmetry see \cite{Kapustin:2006ic}.}

A generalized K\"{a}hler structure on a manifold $\mathcal M$ consists of the triplet $(g,J_\pm,H)$, where $g$ is a Riemannian metric, $J_\pm$ are two integrable complex structures, and $H$ is a closed three-form (that can be locally written as $H=db$), subject to some constraints. This is the most general target space for $\N=(2,2)$ NLSMs, containing K\"{a}hler geometry as the special case $H=0$ and $J_+=\pm J_-$. The complex structures are covariantly constant, $\nabla^{\pm} J_{\pm}=0$, each with respect to a connection with torsion $\nabla^\pm=\nabla^0\pm \tfrac{1}{2} g^{-1}db$, where $\nabla^0$ is the Levi-Civita connection. The presence of torsion implies that the geometry is generically not K\"{a}hler: the forms $\omega_{\pm}=g J_{\pm}$ are not closed. This structure was originally discovered in \cite{Gates:1984nk}, where it was termed bi-Hermitian geometry. More recently, it has been reformulated as the analog of K\"{a}hler geometry in the context of generalized complex geometry \cite{Gualtieri:2003dx, Gualtieri:2010fd}.%
\footnote{For introductory lectures on generalized complex geometry and its relation to supersymmetry, see for instance \cite{Zabzine:2006uz, Koerber:2010bx}.  For a review of generalized K\"{a}hler geometry and general $\N=(2,2)$ NLSMs see \cite{Lindstrom:2005zr}.
}

Locally, on a generalized K\"ahler manifold one can always choose coordinates which are adapted to the decomposition of the tangent bundle
$$
T_{\mathcal M} = \ker (J_+-J_-) \oplus \ker  (J_++J_-) \oplus \coIm  [J_+,J_-] \;,
$$
where the last factor is the co-image \cite{Ivanov:1994ec, Lyakhovich:2002kc}. We denote the  corresponding coordinates by $\Phi \oplus \chi \oplus (\Bbb X_L,\Bbb X_R) $, respectively. In terms of the $\N=(2,2)$ sigma model, these correspond to different matter multiplets: chiral, twisted chiral, left semichiral and right semichiral, respectively. From the perspective of generalized K\"{a}hler geometry, semichiral fields are as fundamental as chiral and twisted chiral: the latter parametrize directions along which the two complex structures commute, while the former parametrize directions along which they do not. As in K\"{a}hler geometry, the full geometric data is locally encoded in a single function: the generalized K\"{a}hler potential $K$ \cite{Lindstrom:2005zr}. Apart from  obeying certain inequalities for the metric to be positive definite, $K=K(\Phi,\bar \Phi;\chi,\bar \chi; \Bbb X_L,\bar{ \Bbb X}_L,\Bbb X_R, \bar{ \Bbb X}_R )$ is otherwise an arbitrary real function of the coordinates on the manifold. As in the K\"{a}hler case, $K$ serves as the action for the NLSM in superspace. The  case of complex dimension three, and in particular the case of one pair of semichiral fields and one chiral field, is especially relevant to supergravity \cite{Halmagyi:2007ft}.

It is shown in \cite{Crichigno:2015pma} that GLSMs with semichiral fields coupled to the usual vector multiplet are continuous deformations of certain GLSMs with chiral fields only, which realize noncompact Calabi-Yau manifolds. The deformation preserves the R-symmetry at the quantum level, but deforms the geometric structure of the target from K\"{a}hler to generalized K\"{a}hler by introducing torsion. Here we place those gauge theories on $S^2$ (with the untwisted background of \cite{Benini:2012ui, Doroud:2012xw}), and compute the exact partition function by supersymmetric localization ``on the Coulomb branch''. It turns out that gauge theories with semichiral fields do not admit enough real masses to lift all massless modes (which appear as non-compact directions in the IR NLSM), therefore their partition function is threatened by divergences. We propose a contour prescription to remove those divergences. We show that the parameters controlling the non-K\"{a}hler deformation enter in a $\Q$-exact term. As a consequence, the $S^2$ partition function should not depend on this deformation and should coincide with the partition function in the K\"{a}hler case. We verify this fact explicitly.

We will also discuss some consequences of our result for topological A/B-models on generalized K\"ahler manifolds. As shown in \cite{Kapustin:2004gv} the topological A-model localizes to generalized holomorphic maps,
\equ{
\label{generalized holo maps}
(1-i J_+) \bar\partial X = 0 \;,\qquad\quad (1 + i J_- ) \partial X = 0 \;,
}
where $X$ are real coordinates on the target $\mathcal M$. For generic $J_\pm$, these equations are very restrictive and the only solutions are constant maps. In our models, the $S^2$ partition function defined with our contour prescription does receive instanton corrections---equal to the ones in the Calabi-Yau before deformation---and yet there are no real compact solutions to (\ref{generalized holo maps}).%
\footnote{We stress that our analysis of the generalized K\"ahler structure is carried out only in the UV of the NLSM, which is enough to capture the instanton corrections. An important question is what the behavior of these theories is in the deep IR. The conditions for conformal invariance at the quantum level, and the relation to the  generalized Calabi-Yau condition of Hitchin \cite{Hitchin:2004ut}, is discussed in \cite{Grisaru:1997pg, Halmagyi:2007ft, Hull:2010sn}. This, however, goes beyond the scope of this paper.}
A possible resolution of this puzzle is that, in fact, our partition function receives contributions from complexified solutions to (\ref{generalized holo maps}), which are less restricted and may well exist even for generic $J_\pm$. We leave this puzzle as an open question.

The paper is organized as follows. In Section~\ref{section semichiral fields on R2} we review the field components, supersymmetry transformations, gauged supersymmetric actions for semichiral fields on $\Bbb R^{2}$ and discuss the NLSMs these theories realize. In Section~\ref{section semichiral fields on S2} we place these gauge theories on  $S^{2}$. In Section~\ref{Localization on the Coulomb Branch} we study the BPS configurations, localize the path-integral, and discuss issues of the integration contour and instanton contributions. In Section~\ref{Sigma Models on Generalized Kahler Manifolds} we review some aspects of topologically twisted NLSMs on generalized K\"{a}hler manifolds, and study the phenomenon of type-change in an example of the resolved conifold with a generalized K\"{a}hler structure. We conclude with a discussion in Section~\ref{Discussion}.

\section[\texorpdfstring{Semichiral Fields on $\bR^2$}{Semichiral Fields on R²}]{Semichiral Fields on $\boldsymbol{\pmb{\bR}^2}$}\label{section semichiral fields on R2}

We begin by reviewing some basic aspects of $\N=(2,2)$ supersymmetry and defining our notation and conventions. We then discuss gauge theories for semichiral fields and their description as NLSMs. 

The algebra of $\N=(2,2)$ superderivatives is
\equ{
\label{algebra spinor derivatives ungauged}
\{ \Bbb D_\pm, \bar{\Bbb D}_\pm \} = \pm2i \partial_{\pm\pm} \;,
}
where $\pm$ are spinor indices, ${\Bbb D_\pm}, \bar{\Bbb D}_\pm$ are superderivatives and $\partial_{\pm\pm} = \frac{1}{2}(\partial_1 \mp i \partial_2)$ are spacetime derivatives; the precise definitions are given in Appendix \ref{app: susy}. The SUSY transformations are generated by
\equ{ 
\delta = \bar{ \epsilon}^{+} \Bbb Q_{+} + \bar{\epsilon}^{-} \Bbb Q_{-}+ \epsilon^{+} \bar{\Bbb Q}_{+}+ \epsilon^{-} \bar{\Bbb Q}_{-} \;,
}
where $\epsilon$, $\bar \epsilon$ are anticommuting Dirac spinors while $\bQ$, $\bar\bQ$ are the supercharges satisfying $\{\Bbb Q_{\pm},\bar{\Bbb Q}_{\pm}\}=\mp 2i \partial_{\pm\pm}$ and anticommuting with the spinor derivatives: $\{\Bbb Q_{\pm},\Bbb D_{\pm}\}=0$, \etc

\subsection{Supermultiplets}

The basic matter supermultiplets are chiral, twisted chiral and semichiral fields. In \textit{Lorentzian} signature these fields are defined by the following set of constraints:
\equ{
\label{constraints multiplets}
\begin{array}{rllll}
\text{Chiral}: &\bar{ \bD}_{+}\Phi=0 \;,\quad &\bar{ \bD}_{-}\Phi=0 \;,\quad &\bD_{+}\bar \Phi=0 \;,\quad &\bD_{-} \bar \Phi=0 \;, \\ 
\text{Twisted Chiral}: &\bar{ \bD}_{+}\chi=0 \;,\quad  &\bD_{-}\chi=0 \;,\quad &\bD_{+}\bar \chi=0 \;,\quad &\bar{\bD}_{-} \bar \chi=0 \;, \\  
\text{Left semichiral}: &\bar{ \bD}_{+}\Bbb X_{L}=0 \;,\quad & & \bD_{+}\bar{\Bbb X}_{L}=0 \;, \\
\text{Right semichiral}: & & \bar{ \bD}_{-}\Bbb X_{R}=0 \;,\quad & & \bD_{-}\bar{\Bbb X}_{R}=0 \;.
\end{array}
}
In Lorentzian signature, complex conjugation acts on superderivatives as $\bD_\pm^\dag = \bar{\bD}_\pm$ and on superfields as $\Bbb X^\dag = \bar{\Bbb X}$. The SUSY constraints (\ref{constraints multiplets}) are compatible with complex conjugation.

In \textit{Euclidean} signature, however, the conjugation of superderivatives changes the helicity, namely $\bD_{\pm}^\dag= \bar{\bD}_{\mp}$, and taking the complex conjugate of the constraints (\ref{constraints multiplets}) may lead to additional constraints. In the case of a twisted chiral field $\chi$, for instance, this implies that the field be constant. The well-known resolution is to complexify the multiplet and consider $\chi$ and $\bar{\chi}$ as independent fields. Although this problem does not arise for semichiral fields, we nonetheless choose to complexify them.%
\footnote{One may choose not to do so. Then, a left semichiral field $\Bbb Y_L$ satisfies $\bar{\bD}_{+} \Bbb Y_{L}=0$ and its (Euclidean) Hermitian conjugate $\bar{\Bbb Y}_L$ satisfies $\bD_{-} \bar{\Bbb Y}_{L}=0$, and similarly for a right semichiral field. However, the target space geometry of these models is not well understood. Since ultimately we are interested in learning about  the target space geometry of models in Lorentzian signature, we choose to complexify semichiral fields.}
That is, we will consider $\Bbb X_L$ a left semichiral field and $\bar{\Bbb X}_L$ an independent left anti-semichiral field, and similarly for $\Bbb X_R$ and $\bar{\Bbb X}_R$. The SUSY constraints (and their Euclidean conjugates) read:
\equ{
\begin{array}{rrrr}
\bar{ \bD}_{+}\Bbb X_{L} = 0 \;,\quad & \bD_{+}\bar{\Bbb X}_{L} = 0 \;,\quad & \bar{ \bD}_{-}\Bbb X_{R}=0 \;,\quad & \bD_{-}\bar{\Bbb X}_{R}=0 \;, \\
 \bD_{-}\Bbb X_{L}^{\dagger}=0 \;,\quad & \bar{\bD}_- \bar{\Bbb X}_{L}^{\dagger}=0 \;,\quad  & \bD_{+}\Bbb X_{R}^{\dagger}=0 \;,\quad & \bar{\bD}_{+}\bar{\Bbb X}_{R}^{\dagger}=0 \;.
\end{array}
}
The target space geometry of these models is the complexification of the target space geometry of the corresponding models defined in Lorentzian signature. See \cite{Hull:2008de} for a discussion of these issues.

\subsection{Components and  Supersymmetry Transformations}

Semichiral fields were originally introduced in \cite{Buscher:1987uw}. Since they are less known than chiral and twisted chiral fields, we review some of their basic properties here. Each left or right semichiral multiplet consists of 3 complex scalars, 4 Weyl fermions, and one complex chiral vector. We denote these by
\bea
\label{components left and right semi}
\Bbb X_{L}:\,&\quad (X_{L},\psi_{\pm}^{L},F_{L}, \bar \chi_{-},M_{-+},M_{--},\bar \eta_{-}) \;, \qquad \bar{\Bbb X}_L :\,\quad ( \bar X_L, \bar\psi_\pm^L, \bar F_L, \chi_-, \bar M_{-+}, \bar M_{--}, \eta_-) \;,\\
\Bbb X_{R}:\,&\quad (X_{R},\psi_{\pm}^{R},F_{R}, \bar \chi_{+},M_{+-},M_{++},\bar \eta_{+}) \;, \qquad \bar{\Bbb X}_R :\,\quad (\bar X_R, \bar\psi_\pm^R, \bar F_R, \chi_+, \bar M_{+-}, \bar M_{++}, \eta_+)\,,
\eea
where $\psi_\alpha, \chi_\alpha, \eta_\alpha$ are fermionic and $X,F,M_{\alpha \beta}$ are bosonic fields, all valued in the same representation $\fR$ of a gauge group $G$ (see Appendix~\ref{N=(2,2) Multiplets} for details). Compared to the field content of chiral and twisted chiral fields, semichiral fields have twice the number of bosonic and fermionic components.

To treat left and right semichiral fields in a unified way, we define a superfield $\Bbb X$ that satisfies at least one chiral constraint (either $\bar \bD_{+} \Bbb X=0$ or $\bar \bD_{-} \Bbb X=0$, or both), but we do not specify which one until the end of the calculation. Similarly, $\bar{\Bbb X}$ is an independent field satisfying at least one antichiral constraint (either $\bD_{+} \bar{ \Bbb X}=0$ or $ \bD_{-} \bar{\Bbb X}=0$, or both). We denote the field content by
\equ{
\label{definitions component X ch2}
\Bbb X:\quad (X,\psi_{\alpha},F, \bar \chi_{\alpha},M_{\alpha \beta},\bar \eta_{\alpha})\,,  \qquad \text{with}\quad \alpha, \beta=\pm \;,
}
and similarly for $\bar{\Bbb X}$. By setting $\bar \chi_\alpha = M_{\alpha\beta} = \bar\eta_\alpha = 0$, the multiplet $\Bbb X$ describes a chiral multiplet $\Phi: (X, \psi_\alpha, F)$. By setting $\bar \chi_{+}=M_{+-}=M_{++}=\bar{\eta}_{+}=0$ it describes a left semichiral multiplet, and by setting $\bar \chi_- = M_{-+} = M_{--} = \bar\eta_- = 0$ it describes a right semichiral field.\footnote{With appropriate identifications, it can also describe a twisted chiral multiplet. However, since we are interested in minimally coupling $\Bbb X$ to the usual vector multiplet, we do not consider that case here.} 

We couple the multiplet $\Bbb X$  to the usual vector multiplet $V$. One could also consider couplings to other vector multiplets, but we do not do that here (see instead \cite{Crichigno:2015pma}). The field content of the usual vector multiplet is $(A_\mu,\sigma_1, \sigma_2, \lambda_\pm, D)$, where $\sigma_1,\sigma_2$ are real in Lorentzian but complex in Euclidean signature and we define
\equ{
\sigma = i\sigma_1 - \sigma_2 \;,\qquad\quad \bar\sigma = -i \sigma_1 - \sigma_2 \;.
}
The SUSY transformation rules for the multiplet $\Bbb X$, minimally coupled to the vector multiplet $V$, are derived in Appendix~\ref{N=(2,2) Multiplets} and read:
\bea
\label{transf rules flat space}
\delta \psi_\alpha &= \big(\!\left[ i \gamma^\mu D_\mu X + i\sigma_1 X + \sigma_2 X \gamma_3 \right] \epsilon \big)_\alpha - \epsilon^\beta M_{\beta \alpha} + \bar{\epsilon}_\alpha F \qquad &
\delta X &= \bar{\epsilon} \psi + \epsilon \bar{\chi} \\
\delta F & = \left[-i\sigma_1 \psi - \sigma_2 \psi \gamma_3 - i\lambda X - i (D_\mu \psi) \gamma^\mu  + \bar{\eta} \right] \epsilon \qquad &
\delta \bar\chi_\alpha & = \bar{\epsilon}^\beta M_{\alpha\beta} \\
\delta M_{\alpha\beta} & = -\bar{\eta}_\alpha \bar{\epsilon}_\beta - i \sigma_1 \bar{\chi}_\alpha \epsilon_\beta - \sigma_2 \bar{\chi}_\alpha \gamma_3 \epsilon_\beta - i(D_\mu \bar{\chi}_\alpha) (\gamma^\mu \epsilon)_\beta \qquad \\
\delta \bar{\eta}_\alpha & = -i (\epsilon \lambda) \bar{\chi}_\alpha + i(\epsilon \gamma^\mu)^\beta D_\mu M_{\alpha \beta} - i \sigma_1 \epsilon^\beta M_{\alpha\beta} - \sigma_2 (\gamma_3 \epsilon)^\beta M_{\alpha\beta} \;,
\eea
and similarly for the multiplet $\bar{\Bbb X}$:
\bea
\delta \bar{\psi}_\alpha &= \big(\!\left[ i \gamma^\mu D_\mu \bar{X} - i\sigma_1 \bar{X} + \sigma_2 \bar{X} \gamma_3  \right] \bar{\epsilon} \big)_\alpha - \bar{\epsilon}^\beta \bar{M}_{\beta\alpha} + \epsilon_\alpha \bar{F} \qquad &
\delta \bar{X} &= \epsilon \bar{\psi} + \bar{\epsilon} \chi \\
\delta \bar{F} & = \left[i\sigma_1 \bar{\psi} - \sigma_2 \bar{\psi} \gamma_3 - i\bar{\lambda} \bar{X} - i(D_\mu \bar{\psi}) \gamma^\mu + \eta \right] \bar{\epsilon} \qquad &
\delta \chi_\alpha &= \epsilon^\beta \bar{M}_{\alpha \beta} \\
\delta \bar{M}_{\alpha\beta} & = -\eta_{\alpha} \epsilon_{\beta} + i\sigma_1 \chi_\alpha \bar{\epsilon}_\beta - \sigma_2 \chi_\alpha \gamma_3 \bar{\epsilon}_\beta - i(D_\mu \chi_\alpha) (\gamma^\mu \bar{\epsilon})_\beta \qquad \\
\delta \eta_\alpha & = -i (\bar{\epsilon} \bar{\lambda}) \chi_\alpha + i (\bar{\epsilon} \gamma^\mu)^\beta D_\mu \bar{M}_{\alpha \beta} + i\sigma_1 \bar{\epsilon}^\beta \bar{M}_{\alpha\beta} - \sigma_2 (\gamma_3 \bar{\epsilon})^\beta \bar{M}_{\alpha \beta} \;.
\eea
Here $D_\mu=\partial_\mu-i A_{\mu}$ is the gauge-covariant derivative. To keep the notation compact in what follows, it is convenient to introduce the operator
\equn{
\P_{\alpha\beta} \,\equiv\, \pmat{2i D_{++} & \sigma \\ \bar\sigma & -2i D_{--}} \;.
}

\subsection{Supersymmetric Actions}

The gauge-invariant kinetic action for semichiral fields in flat space is built out of terms of the form:
\eqs{
\label{action general flat space}
\L_{\Bbb X}^{\Bbb R^{2}} &= \int d^{4} \theta \, \bar{\Bbb X} \, \Bbb X\nonumber \\
&= D_{\mu}\bar{X}D^{\mu} X + \bar{X}\(\sigma_1^{2}+\sigma_2^{2} + i D \)X +\bar{F}F - \bar{M}_{\alpha\beta}M^{\beta\alpha} -\bar{X}\P_{\alpha\beta}M^{\alpha\beta} +\bar{M}^{\alpha\beta}\P_{\beta\alpha}X \nonumber\\
&\quad -\bar{\psi} \( i \gamma^{\mu}D_{\mu} -i\sigma_1 + \gamma_3 \sigma_2 \) \psi + i \bar{\psi}\lambda X - i \bar{X}\bar{\lambda}\psi - \eta\psi - \bar{\psi}\bar{\eta} \nonumber\\  
&\quad + \bar{\chi} \( i \gamma^{\mu}D_{\mu} - i \sigma_1 + \gamma_3 \sigma_2 \) \chi + i \bar{X}\lambda\bar{\chi} - i\chi\bar{\lambda}X \;.
}
Setting some fields to zero, according to the  discussion below (\ref{definitions component X ch2}), gives the corresponding action for chiral and semichiral fields. An important aspect of semichiral fields, however, is that in order to obtain standard quadratic kinetic terms after having integrated out the auxiliary fields, they must come in pairs $(\Bbb X_L, \Bbb X_R)$. Consider for instance a (neutral, for simplicity) left semichiral field with kinetic action 
\equn{
\int d^{4} \theta \, \bar{\Bbb X}_L \, \Bbb X_L \;.
}
It is easy to see that the equations of motion set $M_{-+}=\bar M_{-+}=\psi_+=\bar \psi_+=0$ and give the first order equations
\eqsn{
\partial_{++}X_L = \;& \partial_{++}\bar{X}_L=\partial_{++}M_{--}=\partial_{++}\bar M_{--}=0 \\
\partial_{++}\psi_- = \;&\partial_{++}\bar \psi_-= \partial_{++}\chi_-=\partial_{++}\bar \chi_-=0 \;,
}
which describe two left-moving bosonic and two left-moving fermionic modes. Although interesting, we leave the study of such Lagrangians for future work.

To obtain sigma models with standard kinetic terms, one must consider models with the same number of left and right semichiral fields and an appropriate coupling between them, either of the form $(\bar{\Bbb X}_L \Bbb X_R+c.c.)$ or $(\Bbb X_L \Bbb X_R+c.c.)$. In such models, integrating out the auxiliary fields leads to standard kinetic terms as well as (gauged) Wess-Zumino couplings. Note that depending on the type of the off-diagonal term one chooses, left and right semichiral fields must be either in the same or in conjugate representations of the gauge group.  Thus, from now on we restrict ourselves to models containing pairs of semichiral fields $(\Bbb X_{L}, \Bbb X_{R})$ either in a representation $(\fR, \fR)$ or $(\fR, \overline\fR)$ of the gauge group.

\subsection*{\it Same Representation}

Consider a pair of semichiral fields $(\Bbb X_L, \Bbb X_R)$ in representation $(\fR, \fR)$. The most general gauge-invariant quadratic action follows from the Lagrangian
\equ{
\label{flat space lagrangian left and right}
\L_{LR}^{\Bbb R^{2}}=-\int d^{4}\theta \left[\bar{\Bbb X}_{L}\Bbb X_{L}  +\bar{\Bbb X}_{R} \Bbb X_{R} +\alpha \big( \bar{\Bbb X}_{L} \Bbb X_{R}  + \bar{\Bbb X}_{R}  \Bbb X_{L} \big) \right] \;,
}
where $\alpha>1$ is a real parameter.%
\footnote{One may begin with the general  Lagrangian 
$$
\L = \int d^4\theta\, \left[\beta \, \bar{\Bbb X}_{L} \Bbb X_{L} + \gamma \, \bar{\Bbb X}_{R} \Bbb X_{R} - \alpha \, \bar{\Bbb X}_{L} \Bbb X_{R} - \alpha^* \, \bar{\Bbb X}_{R} \Bbb X_{L} \right]\,
$$
where $\beta$, $\gamma$ are real parameters and $\alpha$ is complex. This describes flat space and requiring that the metric be positive requires $\beta,\gamma \neq 0$. Then, by rescaling the fields one can set $\beta=\pm1$ and $\gamma=\pm1$. By a further phase redefinition of the fields, $\alpha$ can be made real and non-negative. Finally, the requirement of a positive-definite metric implies $\beta = \gamma = -1$, $\alpha>1$.
}
Before gauging, the action (\ref{flat space lagrangian left and right}) describes flat space with a constant $B$-field controlled by $\alpha$, which of course could be gauged away; the significance of the parameter $\alpha$ is that it determines a choice of complex structures $J_\pm$ on the space.\footnote{See \cite{Crichigno:2015pma} and references therein for a more detailed discussion on this point.} We note that due to the the  off-diagonal term, $\Bbb X_L$ and $\Bbb X_R$ must have the same R-charge.

The general case with multiple pairs of semichiral fields is very similar. Under the assumption that the metric is positive definite, one can always diagonalize the kinetic term and reduce it to multiple decoupled copies of \eqref{flat space lagrangian left and right} by field redefinitions; see Appendix~\ref{Kinetic action and positivity of the metric} for details.

To gain some intuition on GLSMs for semichiral fields, and their difference with standard GLSMs for chiral fields, we reduce the Lagrangian (\ref{flat space lagrangian left and right})  to component fields and integrate out the auxiliary components in the semichiral multiplet. This gives
\equ{
\label{action components no auxiliaries}
\L_{LR}^{\Bbb R^{2}}= \big( g_{mn} + b_{mn} \big) \, D_{++} X^m D_{--} X^n + \frac{\alpha}{2} \Big( \bar X \big( |\sigma|^2 + D \big) X + \bar{\tilde X} \big( |\sigma|^2 + D \big) \tilde X \Big) + \ldots  \;,
}
where we have omitted fermionic kinetic terms and Yukawa couplings, and defined
\equ{
\label{flat metric and B-field}
g_{mn} = \pmat{
0 &  1 & 0 & 0 \\ 
1 &  0 & 0 & 0 \\ 
0 &  0 & 0 & 1 \\ 
0 &  0 & 1 & 0 } ,\qquad\qquad
b_{mn}=\sqrt{\alpha^2-1} \pmat{
0 &  0 & 1 & 0 \\ 
0 &  0 & 0 & 1 \\ 
-1 &  0 & 0 & 0 \\ 
0 &  -1 & 0 & 0 } ,
}
where we denote $X^m=(X,\bar X, \tilde X, \bar{ \tilde X})$ and we made the change of variables
\equ{
X_L = \sqrt{\tfrac{\alpha}{4(\alpha+1)}}\, X+\sqrt{\tfrac{\alpha}{4(\alpha-1)}} \, \bar{\tilde X} \;,\qquad\quad X_R = \sqrt{\tfrac{\alpha}{4(\alpha+1)}}\, X-\sqrt{\tfrac{\alpha}{4(\alpha-1)}} \, \bar{\tilde X} \;.
}
We note that after this field redefinition one may take the special limit $\alpha \rightarrow 1$, for which $b_{mn}\to 0$. In fact, in this limit the whole Lagrangian (\ref{action components no auxiliaries}) coincides with the component Lagrangian of a standard GLSM for chiral multiplets (with lowest components $X$, $\tilde X$) in gauge representations $(\fR, \overline \fR)$. Thus, the parameter $\alpha$ controls a deformation of such model  which includes the addition of a gauged Wess-Zumino term. The deformation is not generic: it is such that by adding suitable auxiliary fields $\N=(2,2)$ SUSY can be realized off-shell in terms of semichiral fields.%
\footnote{For a discussion of more general gauge theories with gauged Wess-Zumino terms, but only \textit{on-shell} $\N=(2,2)$ SUSY, see \cite{Kapustin:2006ic}.} 

As we shall discuss in Section~\ref{NLSM description}, these models give rise to NLSMs on  generalized K\"{a}hler manifolds with non-zero three-form $H$, controlled by the parameter $\alpha$. In the limit $\alpha \to 1$, the $H$ field vanishes and $g$ becomes a K\"{a}hler metric, as expected.

\subsection*{\it Conjugate Representations}

For a pair of semichiral fields $(\Bbb X_L,\Bbb X_R)$ in conjugate representations $(\fR, \wb\fR)$, the most general gauge-invariant quadratic Lagrangian is
\equ{
\label{flat space lagrangian left and right opposite charges}
\L_{LR}^{\Bbb R^{2}}=\int d^{4}\theta \left[\bar{\Bbb X}_{L} \Bbb X_{L}  +\bar{\Bbb X}_{R}  \Bbb X_{R} +\beta \big( \Bbb X_{L}  \Bbb X_{R}  + \bar{\Bbb X}_{R} \bar{ \Bbb X}_{L} \big) \right] \;,
}
where $\beta>1$ is a real parameter.
In this model $\Bbb X_L$ and $\Bbb X_R$ have opposite R-charges. Once again, reducing the Lagrangian to components and integrating out the auxiliary fields, one observes that the theory is a deformation of a standard GLSM with chiral fields in representation $(\fR, \wb\fR)$ by a gauged Wess-Zumino term controlled by the parameter $\beta$, which vanishes in the limit $\beta\rightarrow \infty$. In fact, the gauge theories (\ref{flat space lagrangian left and right}) and (\ref{flat space lagrangian left and right opposite charges}) are related off-shell by a simple field redefinition corresponding to a change of coordinates on the target space and sending $\alpha \to 1$ corresponds to sending $\beta \to \infty$ (see Appendix~\ref{Semichiral-Semichiral Duality}). Thus, without loss of generality, we shall consider only one of these actions, choosing the most convenient one for a particular calculation. In the multiflavor case, by field redefinitions, one can always rewrite the Lagrangian as multiple decoupled copies of (\ref{flat space lagrangian left and right opposite charges}).

Finally, we comment that traditional superpotential terms for semichiral fields are not possible because they break supersymmetry. Fermionic superpotential terms (integrals over $d^{3}\theta $) are possible if a fermionic semichiral multiplet is present, but this is not the case here.

\subsection{NLSM Description}\label{NLSM description}

As mentioned above, one of the main motivations to study gauge theories with semichiral fields is that they realize NLSMs on generalized K\"{a}hler manifolds, as opposed to K\"{a}hler manifolds when only chiral (or only twisted chiral) fields are present. Let us briefly illustrate this point in the case $G=U(1)$, although the discussion holds with arbitrary gauge group.   

Consider $N_{F}$ pairs of semichiral fields $(\Bbb X_{L}^{i},\Bbb X_{R}^{i})$,  $i=1,\ldots,N_{F}$, charged under a $U(1)$ vector multiplet with charges $(Q_{i},-Q_{i})$. The Lagrangian for the GLSM is given by 
\equ{
\label{flat space lagrangian 1,-1 charges}
\L=\L_\text{VM} +\sum_{i=1}^{N_{F}} \int\!\! d^4\theta \left[\bar{\Bbb X}_{L}^{i} \Bbb X_{L}^{i}  +\bar{\Bbb X}_{R}^{i} \Bbb X_{R}^{i} + \beta_i \big( \Bbb X_{L}^{i}  \Bbb X_{R}^{i}  +  \bar{\Bbb X}_{R}^{i} \bar{ \Bbb X}_{L}^{i} \big) \right] + \frac{i}{2} \int\!\! d^2\tilde \theta \, t \, \Sigma + c.c. \;,
}
with $\beta_i>1$, where 
\equ{
\label{action vector multiplet}
\L_\text{VM} = -\frac{1}{2e^2} \int d^4 \theta\, \bar \Sigma \Sigma \;,\qquad \Sigma = \bar{\Bbb D}_+ \Bbb D_- V \;,
}
and $t= i \xi+\tfrac{\theta}{2\pi}$ is the complexified Fayet-Iliopoulos (FI) parameter. In addition, one may have any number of chiral multiplets charged under the same vector multiplet $V$ with arbitrary charges. In the case that only semichiral fields are present, a better formulation of these theories in terms of a new vector multiplet is discussed in \cite{Crichigno:2015pma}. 

Just as in GLSMs with chiral fields \cite{Witten:1993yc}, on the Higgs branch of the theory the vector multiplet becomes massive and at low energies (compared to the gauge coupling $e$) it can be integrated out. The effective theory is a NLSM on the space of vacua, which is determined by the vanishing of the scalar potential $U$, modulo the action of the gauge group. The full moduli space of the GLSMs at hand is analyzed in \cite{Crichigno:2015pma}. For the theory in (\ref{flat space lagrangian 1,-1 charges}), there is a Higgs branch given by%
\footnote{One way to see this is to compute the scalar potential explicitly by going down to components and working, say, in Wess-Zumino gauge. Alternatively, one may work in superspace, by writing $ \Bbb X^{(0)\,i}=  \Bbb X^{i}$ and $\bar{\Bbb X}^{(0)\,i}=\bar{\Bbb X}^{i} e^{-Q_{i}V}$ in (\ref{flat space lagrangian 1,-1 charges})  to introduce the vector multiplet explicitly (here we are following similar notation to that in \cite{Witten:1993yc}). Then, the lowest component of the equation of motion for $V$ leads to the constraint below. Note, in particular, that due to the absence of $e^V$ terms in the off-diagonal terms, the $\beta_{i}$ do not enter in the  constraint; they do however determine the geometric structure on $\mathcal M$. 
}
\equ{
\label{moduli space 1,-1 charges}
\mathcal M = \Big\{ \sum\nolimits_i Q_{i} \Big( \big| X_L^i \big|^2 - \big| X_R^i \big|^2 \Big) - \xi=0 \Big\} \Big/ U(1) \;,
}
where $U(1)$ acts by $X_L^i \to e^{i \alpha Q^i} X_L^i$, $X_R^i \to e^{-i \alpha Q^i} X_R^i$. The complex dimension of $\M$ is $2 N_{F}-1$. Topologically, $\mathcal M$ coincides with the Higgs branch of a GLSM with chiral (as opposed to semichiral) fields $(\Phi^{i},\wt \Phi^{i})$ with gauge charges $(Q_{i},-Q_{i})$.  However, in the semichiral model the space is endowed with a family of generalized K\"{a}hler structures controlled by the parameters $\beta_{i}$. In the limit $\beta_{i}\to \infty$ the space becomes K\"{a}hler, but at finite $\beta_{i}$ there are two independent complex structures $J_{\pm}$ and, due to the presence of semichiral fields, $[J_{+},J_{-}]\neq0$ at generic points on the manifold. We will give an explicit example in Section~\ref{Example: Generalized Conifold}.

One may also consider a model with gauge charges $(Q_i, Q_i)$. As discussed above, this is completely equivalent to the model with charges $(Q_i, -Q_i)$. Thus, the moduli space of these models is always noncompact, for any choice of gauge charges. The generalization of this discussion to generic gauge groups $G$ is straightforward,\footnote{This is certainly the case when there are also charged chiral fields in the model. If only semichiral fields are present, the generalization to the nonabelian case may be more subtle \cite{Crichigno:2015pma}.} at the classical level (the analysis of quantum effects would require a careful treatment, as in \cite{Hori:2006dk}).

It is well known that the moduli spaces (\ref{moduli space 1,-1 charges}) admit not only a K\"{a}hler, but in fact a Calabi-Yau structure (the condition $\sum_a Q_a=0$ is automatically satisfied). It would be interesting to study whether they also admit a generalized Calabi-Yau structure \cite{Hitchin:2004ut}. This could answer the question of the behavior of the theories in the deep IR, which we do not address here. 

Before we proceed we would like to make a comment on couplings to other vector multiplets. Since semichiral fields are less constrained than chiral or twisted chiral fields, they admit minimal couplings to various vector multiplets. In addition to the usual vector multiplet, they may couple minimally to the twisted vector multiplet, as well as to the Semichiral Vector Multiplet (SVM) and the Large Vector Multiplet (LVM) introduced in \cite{Lindstrom:2007vc, Lindstrom:2008hx}. The corresponding GLSMs and the structure of their moduli space  is discussed in \cite{Crichigno:2015pma}. We do not study these models here,  but we comment briefly on the coupling to the Abelian SVM in Section~\ref{Coupling to Other Vector Multiplets}.

\section[\texorpdfstring{Semichiral Fields on $S^2$}{Semichiral Fields on S²}]{Semichiral Fields on $\boldsymbol{S^2}$}\label{section semichiral fields on S2}

The main goal of this section is to place the gauge theories (\ref{flat space lagrangian left and right})---or equivalently (\ref{flat space lagrangian left and right opposite charges})---on the round sphere $S^2$ with no twist, \ie, on the supersymmetric background constructed in \cite{Benini:2012ui, Doroud:2012xw} which preserves four supercharges. Neutral semichiral multiplets, as well as general supersymmetry in two dimensions, have been studied in \cite{Closset:2014pda}. We explicitly construct the supersymmetry variations for gauged semichiral multiplets and their action on $S^2$. As we now show, it turns out that the action is $\Q_A$-exact, therefore the partition function will not depend on the parameters $\alpha$ (or $\beta$) therein. We work in components, rather than in superspace.

\subsection{Supersymmetry Transformations}

One way to determine the supersymmetry transformations on $S^{2}$ is by first constructing the $\N=(2,2)$ superconformal transformations and then specializing to an $SU(2|1)$ sub-algebra. The superconformal transformations can be deduced from the $\N=(2,2)$ super-Poincar\'e transformations (\ref{transf rules flat space}) by covariantizing them with respect to Weyl transformations, as we now review.

Let the scalar component $X$ transform under Weyl transformations with a weight $\frac{q}{2}$, \ie{}, under an infinitesimal Weyl transformation $\delta X = -\frac{q}{2}\Omega\,X$. The supersymmetry transformations (\ref{transf rules flat space}) are not covariant under Weyl transformations, but can be covariantized by adding suitable terms proportional to $\nabla_{\pm} \epsilon$, as explained in \cite{Doroud:2012xw} and reviewed in Appendix~\ref{Weyl Covariance}. Following this procedure, we find that the Weyl-covariant transformations for the superfields $\Bbb X$ and $\bar{\Bbb X}$ are:
\bea
\label{transf rules S2}
\delta \psi_\alpha & = \big(\!\left[i (D_\mu X) \gamma^\mu + i\sigma_1 X + \sigma_2 X \gamma_3 + i \tfrac{q}{2} X \slashed \nabla \right] \epsilon \big)_\alpha - \epsilon^\beta M_{\beta \alpha} + \bar{\epsilon}_\alpha F \quad &
\delta X & = \bar{\epsilon} \psi + \epsilon \bar{\chi} \\
\delta F & = \left[-i\sigma_1 \psi - \sigma_2 \psi \gamma_3 - i\lambda X - i (D_\mu \psi) \gamma^\mu - i\tfrac{q}{2} \psi \slashed \nabla + \bar{\eta} \right] \epsilon &
\delta \bar{\chi}_\alpha & = \bar{\epsilon}^\beta M_{\alpha\beta} \\
\delta M_{\alpha\beta} & = -\bar{\eta}_\alpha \bar{\epsilon}_\beta - i \sigma_1 \bar{\chi}_\alpha \epsilon_\beta - \sigma_2 \bar{\chi}_\alpha (\gamma_3 \epsilon)_\beta - i(D_\mu \bar{\chi}_\alpha) (\gamma^\mu \epsilon)_\beta \\
 &\quad - i\tfrac{q+1}{2} \bar{\chi}_\alpha (\slashed \nabla \epsilon)_\beta + \tfrac{i}{2} (\gamma_3 \slashed \nabla \epsilon)_\beta (\gamma_3 \bar{\chi})_\alpha \\
\delta \bar{\eta}_\alpha & = -i (\epsilon \lambda) \bar{\chi}_\alpha + i(\epsilon \gamma^\mu)^\beta D_\mu M_{\alpha \beta} - i \sigma_1 \epsilon^\beta M_{\alpha\beta} - \sigma_2 (\gamma_3 \epsilon)^\beta M_{\alpha\beta} \\
 &\quad + i\tfrac{q+1}{2} \left((\nabla_\mu \epsilon) \gamma^\mu \right)^\beta M_{\alpha \beta} + \tfrac{i}{2} (\nabla_\mu \epsilon \gamma_3 \gamma^\mu)^\beta (\gamma_3)_\alpha\,^\rho M_{\rho\beta} \;,\\
&\hspace*{-1.53cm}\text{and} \\
\delta \bar{\psi}_\alpha & = \big(\!\left[i (D_\mu \bar X) \gamma^\mu - i\sigma_1 \bar{X} + \sigma_2 \bar{X} \gamma_3 + i\tfrac{q}{2} \bar X \slashed \nabla \right] \bar{\epsilon} \big)_\alpha - \bar{\epsilon}^\beta \bar M_{\beta\alpha} + \epsilon_\alpha \bar{F} \quad &
\delta \bar{X} & = \epsilon \bar{\psi} + \bar{\epsilon} \chi \\
\delta \bar{F} & = \left[i\sigma_1 \bar{\psi} - \sigma_2 \bar{\psi} \gamma_3 - i \bar\lambda \bar X - i(D_\mu \bar\psi) \gamma^\mu - i \tfrac{q}{2} \bar{\psi} \slashed \nabla + \eta \right] \bar{\epsilon} &
\delta \chi_\alpha & = \epsilon^\beta \bar{M}_{\alpha \beta} \\
 \delta \bar{M}_{\alpha\beta} & = -\eta_{\alpha} \epsilon_{\beta} + i\sigma_1 \chi_\alpha \bar{\epsilon}_\beta - \sigma_2 \chi_\alpha (\gamma_3 \bar{\epsilon})_\beta - i(D_\mu \chi_\alpha) (\gamma^\mu \bar{\epsilon})_\beta \\
 &\quad - i\tfrac{q+1}{2} \chi_\alpha (\slashed \nabla \bar{\epsilon})_\beta + \tfrac{i}{2} (\gamma_3 \slashed \nabla \bar{\epsilon})_\beta (\gamma_3 \chi)_\alpha \\
 \delta \eta_\alpha & = -i (\bar{\epsilon} \bar{\lambda}) \chi_\alpha + i (\bar{\epsilon} \gamma^\mu)^\beta D_\mu \bar{M}_{\alpha \beta} + i\sigma_1 \bar{\epsilon}^\beta \bar{M}_{\alpha\beta} - \sigma_2 (\gamma_3 \bar{\epsilon})^\beta \bar{M}_{\alpha \beta} \\
 &\quad + i\tfrac{q+1}{2} \left((\nabla_\mu \bar{\epsilon}) \gamma^\mu \right)^\beta \bar{M}_{\alpha\beta} + \tfrac{i}{2} (\nabla_\mu \bar{\epsilon} \gamma_3 \gamma^\mu)^\beta (\gamma_3)_\alpha\,^\rho \bar{M}_{\rho\beta} \;.
\eea
Here $D_\mu=\nabla_\mu-i A_{\mu}$ is the gauge-covariant derivative on $S^2$. Splitting $\delta=\delta_{\e}+\delta_{\bar \e}$ and imposing the Killing spinor equations $\nabla_\mu \epsilon = \gamma_\mu \check\epsilon$, $\nabla_\mu \bar\epsilon = \gamma_\mu \check{\bar\epsilon}$ for some spinors $\check\epsilon$, $\check{\bar\epsilon}$, one finds that the superconformal algebra is realized on semichiral fields as:
\eqsg{\label{algebra}
[\delta_\epsilon,\, \delta_{\bar{\epsilon}}] X & = \xi^\mu \partial_\mu X + i \Lambda X + \tfrac{q}{2} \rho X + iq\alpha X \,, \\
[\delta_\epsilon,\, \delta_{\bar{\epsilon}}] \bar{X} & = \xi^\mu \partial_\mu \bar{X} - i \Lambda \bar{X} + \tfrac{q}{2} \rho \bar{X} - iq\alpha \bar{X} \,, \\
[\delta_\epsilon,\, \delta_{\bar{\epsilon}}] \psi & = \xi^\mu \partial_\mu \psi + i\Lambda \psi + \tfrac{q+1}{2} \rho \psi + i(q-1) \alpha \psi + \tfrac{1}{4} \Theta^{\mu\nu} \gamma_{\mu\nu} \psi + i\beta \gamma_3 \psi \,, \\
[\delta_\epsilon,\, \delta_{\bar{\epsilon}}] \bar{\psi} & = \xi^\mu \partial_\mu \bar{\psi} - i\Lambda \bar{\psi} + \tfrac{q+1}{2} \rho \bar{\psi} - i(q-1) \alpha \bar{\psi} + \tfrac{1}{4} \Theta^{\mu\nu} \gamma_{\mu\nu} \bar{\psi} - i\beta \gamma_3 \bar{\psi} \,, \\
[\delta_\epsilon,\, \delta_{\bar{\epsilon}}] F & = \xi^\mu \partial_\mu F + i\Lambda F + \tfrac{q+2}{2} \rho F + i(q-2) \alpha F \,, \\
[\delta_\epsilon,\, \delta_{\bar{\epsilon}}] \bar{F} & = \xi^\mu \partial_\mu \bar{F} - i\Lambda \bar{F} + \tfrac{q+2}{2} \rho \bar{F} - i(q-2) \alpha \bar{F} \,, \\
[\delta_\epsilon,\, \delta_{\bar{\epsilon}}] \bar{\chi}_\alpha & = \xi^\mu \partial_\mu \bar{\chi}_\alpha + i \Lambda \bar{\chi}_\alpha + \tfrac{q+1}{2} \rho \bar{\chi}_\alpha + i(q+1)\alpha \bar{\chi}_\alpha + \tfrac{1}{4} \Theta^{\mu\nu} \gamma_{\mu\nu} \bar{\chi}_\alpha - i\beta (\gamma_3 \bar{\chi})_\alpha\, ,\\
[\delta_\epsilon,\, \delta_{\bar{\epsilon}}] \chi_\alpha & = \xi^\mu \partial_\mu \chi_\alpha - i \Lambda \chi_\alpha + \tfrac{q+1}{2} \rho \chi_\alpha - i(q+1)\alpha \chi_\alpha + \tfrac{1}{4} \Theta^{\mu\nu} \gamma_{\mu\nu} \chi_\alpha + i\beta (\gamma_3 \chi)_\alpha \,, \\
[\delta_\epsilon,\, \delta_{\bar{\epsilon}}] M_{\alpha\beta} & = \xi^\mu \partial_\mu M_{\alpha\beta} + i\Lambda M_{\alpha\beta} + \tfrac{q+2}{2} \rho M_{\alpha\beta} + iq \alpha M_{\alpha\beta} + \tfrac{1}{4} \Theta^{\mu\nu} (\gamma_{\mu\nu})_\alpha\,^\rho M_{\rho\beta} \\
 &\quad + \tfrac{1}{4} \Theta^{\mu\nu} (\gamma_{\mu\nu})_\beta\,^\rho M_{\alpha\rho} -i\beta (\gamma_3)_\alpha\,^\rho M_{\rho\beta} + i\beta (\gamma_3)_\beta\,^\rho M_{\alpha\rho} \,, \\
[\delta_\epsilon,\, \delta_{\bar{\epsilon}}] \bar{M}_{\alpha\beta} & = \xi^\mu \partial_\mu \bar{M}_{\alpha\beta} - i\Lambda \bar{M}_{\alpha\beta} + \tfrac{q+2}{2} \rho \bar{M}_{\alpha\beta} - iq \alpha \bar{M}_{\alpha\beta} + \tfrac{1}{4} \Theta^{\mu\nu} (\gamma_{\mu\nu})_\alpha\,^\rho \bar{M}_{\rho\beta} \\
 &\quad + \tfrac{1}{4} \Theta^{\mu\nu} (\gamma_{\mu\nu})_\beta\,^\rho \bar{M}_{\alpha\rho} + i\beta (\gamma_3)_\alpha\,^\rho \bar{M}_{\rho\beta} - i\beta (\gamma_3)_\beta\,^\rho \bar{M}_{\alpha\rho} \,, \\
[\delta_\epsilon,\, \delta_{\bar{\epsilon}}] \bar{\eta}_\alpha & = \xi^\mu \partial_\mu \bar{\eta}_\alpha + i\Lambda \bar{\eta}_\alpha + \tfrac{q+3}{2} \rho \bar{\eta}_\alpha + i(q-1) \alpha \bar{\eta}_\alpha + \tfrac{1}{4} \Theta^{\mu\nu} \gamma_{\mu\nu} \bar{\eta}_\alpha - i\beta (\gamma_3 \bar{\eta})_\alpha \,, \\
[\delta_\epsilon,\, \delta_{\bar{\epsilon}}] \eta_\alpha & = \xi^\mu \partial_\mu \eta_\alpha - i\Lambda \eta_\alpha + \tfrac{q+3}{2} \rho \eta_\alpha - i(q-1) \alpha \eta_\alpha + \tfrac{1}{4} \Theta^{\mu\nu} \gamma_{\mu\nu} \eta_\alpha + i\beta (\gamma_3 \eta)_\alpha \, ,
}
where the parameters are given by 
\eqsn{\begin{split}
\xi_\mu  \equiv\,& i(\bar{\epsilon} \gamma_\mu \epsilon)\,,\\
\Theta^{\mu\nu}  \equiv\,& D^{[\mu} \xi^{\nu]} + \xi^\rho \omega_\rho\,^{\mu\nu}\,,\\
\alpha  \equiv\,& -\tfrac{1}{4} (D_\mu \bar{\epsilon} \gamma^\mu \epsilon - \bar{\epsilon} \gamma^\mu D_\mu \epsilon)\,,
\end{split}\qquad
\begin{split}
\Lambda  \equiv\,& -\xi^\mu A_\mu + (\bar{\epsilon} \epsilon) \sigma_1 - i(\bar{\epsilon} \gamma_3 \epsilon) \sigma_2\,,\\
\rho\equiv\,& \tfrac{i}{2} (D_\mu \bar{\epsilon} \gamma^\mu \epsilon + \bar{\epsilon} \gamma^\mu D_\mu \epsilon) = \tfrac{1}{2} D_\mu \xi^\mu\,,\\
\beta\equiv\,& \tfrac{1}{4} (D_\mu \bar{\epsilon} \gamma_3 \gamma^\mu \epsilon - \bar{\epsilon} \gamma_3 \gamma^\mu D_\mu \epsilon)\,,
\end{split}
}
and $\omega_\rho\,^{\mu\nu}$ is the spin connection. Here $\xi_\mu$ parametrizes translations, $Λ$ is a gauge parameter, $\rho$ is a parameter for dilations, and $\alpha, \beta$ parametrize vector and axial R-symmetry transformations, respectively. From here one reads off the charges of the fields under these transformations, which  are summarized in Table~\ref{all charges components semi}. Note that the vector R-charge of the multiplet is twice its Weyl weight, as for chiral fields.  All other commutators vanish, $[\delta_{\e},\delta_{\e}]=[\delta_{\bep},\delta_{\bep}]=0$, if one imposes the extra condition $\Box \e= h \e$, $\Box \bar\epsilon = h\bar\epsilon$ with the same function $h$.

\tabl{h!}{
\begin{tabular}{|c|c|c|c|c|c|c|c|c|c|c|c|c|}
\hline
 & $\Bbb D_{+}$&  $\Bbb D_{-}$& $\bar{\Bbb D}_{+}$& $\bar{\Bbb D}_{-}$&$X$ &  $\psi$ &  $\bar{ \chi}$  & $F$ & $M_{\alpha \beta}$ &  $\bar{\eta}$\\
\hline 
$w$& $\frac{1}{2}$& $\frac{1}{2}$ & $\frac{1}{2}$& $\frac{1}{2}$&$\frac{q}{2}$  &$\frac{q+1}{2}$  &$\frac{q+1}{2}$ &$\frac{q+2}{2}$  &$\frac{q+2}{2}$  &$\frac{q+3}{2}$ \\
\hline
$q_V$ &$ -1$&$-1$ &$1$ & $1$ & $q$  & $q-1$ &  $q+1$ &  $q-2$ & $q$ & $q-1$ \\
\hline
$q_A$  & $1$& $-1$ &$-1$ &$1$ & $0$  & $1$  & $-1$ & $0$ & $-2 \e_{\alpha \beta}$ & $-1$  \\
\hline
\end{tabular}
\caption{Weyl weight, vector and axial R-charge for the component fields of the semichiral multiplet.}
\label{all charges components semi}
}

There are four complex Killing spinors on $S^2$ satisfying $\nabla_\mu \epsilon = \pm \frac i{2r} \gamma_\mu \epsilon$. Restricting the transformations (\ref{transf rules S2}) to spinors $\epsilon, \bar\epsilon$ that satisfy
\equ{
\label{Positive Killing Spinors}
\nabla_{\mu}\e = \tfrac{i}{2r} \gamma_{\mu}\e \;,\qquad\quad \nabla_{\mu}\bep = \tfrac{i}{2r} \gamma_{\mu}\bep \;,
}
the algebra (\ref{algebra}) does not contain dilations nor axial R-rotations (\ie{} $\rho = \beta = 0$). This is an $SU(2|1)$ subalgebra of the superconformal algebra that we identify as the $\N=(2,2)$ SUSY on $S^2$ and denote it by $SU(2|1)_A$, as in \cite{Benini:2012ui, Doroud:2012xw}. The transformation rules in \eqref{transf rules S2} simplify to
\bea
\label{SUSY variations sphere 1}
\delta \psi_{\alpha} &= \epsilon^\beta( \P_{\beta \alpha} X-M_{\beta\alpha}) +\bep_{\alpha} F -\tfrac{q}{2r}X\epsilon_{\alpha} &\qquad\qquad\qquad
\delta X &=\bar\epsilon \psi + \epsilon \bar\chi \\
\delta F &= \epsilon^\alpha\P_{\alpha \beta} \psi^\beta - i(\epsilon\lambda)X +\e \bar \eta + \tfrac{q}{2r}\epsilon\psi &
\delta \bar\chi_\alpha &= M_{\alpha\beta} \bep^\beta \\
\delta M_{\alpha\beta} &= \epsilon^\gamma\P_{\gamma \beta} \bar{\chi}_\alpha - \bar{\eta}_\alpha\bar{\epsilon}_\beta + \tfrac{q-2}{2r}\bar{\chi}_{\alpha} \epsilon_\beta +\tfrac{2}{r}\bar{\chi}_{(\alpha}\epsilon_{\beta)} \\
\delta\bar{\eta}_\alpha &= \epsilon^\kappa\P_{\kappa \gamma} M_{\alpha\beta}C^{\gamma\beta} - i(\epsilon \lambda)\bar{\chi}_\alpha+\tfrac{q}{2r}M_{\alpha \beta}\epsilon^\beta +\tfrac{2}{r}M_{[\alpha\beta]}\epsilon^\beta \;,\\
&\hspace*{-1.54cm}\text{and} \\
\delta \bar \psi_{\alpha} &= \bar{\epsilon}^\beta( \P_{\alpha \beta} \bar{X}-\bar{M}_{\beta\alpha}) +\e_{\alpha} \bar F -\tfrac{q}{2r}\bar X \bep_{\alpha} &\qquad
\delta\bar X &= \epsilon \bar\psi + \bar\epsilon \chi \\
\delta \bar F &= \bep^\alpha\P_{\beta\alpha}\bar{\psi}^\beta - i (\bar{\epsilon}\bar{\lambda})\bar{X} +\bar \epsilon \eta\,+\tfrac{q}{2r} \bep \bar{\psi} &\qquad
\delta \chi_\alpha &= \bar M_{\alpha\beta} \e^\beta \\
\delta \bar{M}_{\alpha \beta} &= \bep^{\gamma}\P_{\beta \gamma}\chi_{\alpha} -\eta_{\alpha}\epsilon_{\beta} +\tfrac{q-2}{2r}\chi_{\alpha}\bep_{\beta} +\tfrac{2}{r}\chi_{(\alpha}\bar{\epsilon}_{\beta)} \\
\delta \eta_{\alpha} &= \bep^{\kappa}\P_{\gamma \kappa}\bar{M}_{\alpha \beta}C^{\gamma \beta}- i (\bep \bar \lambda)\chi_{\alpha}+\tfrac{q}{2r} \bar{M}_{\alpha \beta} \bep^\beta +\tfrac{2}{r}\bar{M}_{[\alpha\beta]}\bar{\epsilon}^\beta \;,
\eea
where $C^{\alpha\beta}$ is the antisymmetric tensor with $C^{+-}=1$ and $[αβ]\,,\,(αβ)$ denotes (anti-) symmetrization of indices, respectively.\footnote{Here we have written the transformations using explicit representations for the gamma matrices and properties of spinors (see Appendix~\ref{Spinors in Euclidean space}). We find this convenient for calculations in the upcoming sections.} Another way to derive these transformation rules is by coupling the theory to background supergravity, along the lines of \cite{Festuccia:2011ws}. Using this method, the SUSY transformations (in the case of neutral fields) on more general Riemann surfaces are given in \cite{Closset:2014pda}.

\subsection{Supersymmetric Actions}

The flat-space action (\ref{action general flat space}) is not invariant under the curved-space transformations (\ref{SUSY variations sphere 1}). However, it is possible to add suitable $\frac1r$ and  $\frac1{r^2}$ terms to obtain an invariant Lagrangian:
\eqsg{
\label{action general sphere}
\L^{S^2}_{\Bbb X} &= \L^{\Bbb R^2}_{\Bbb X}+\delta\L \;, \\
\delta\L &= \frac{iq}r\, \bar{X}\sigma_1 X + \frac{q(2-q)}{4r^2} \, \bar{X}X - \frac{q}{2r} \Big( \bar{\psi}\psi + \chi\bar{\chi} + \bar{X} C^{\alpha\beta} M_{\alpha\beta} + C^{\alpha\beta} \bar{M}_{\alpha\beta} X \Big) \;.
}
The first three terms in $\delta\L$ are the ones also needed in the case of a chiral field, while the last three terms are additional ones required for semichiral fields. The action is not only supersymmetric, it is also $\Q_A$-exact, namely:
\equ{
\label{exact action same charges}
\bep \e \int d^{2} x \, \L_{\Bbb X}^{S^{2}} = \delta_{\e}\delta_\bep \int d^{2}x \(\bar \psi \psi + \chi \bar\chi - 2 i \bar{X} \sigma_1 X + \tfrac{q-1}r \, \bar X X+\bar{X}C^{\alpha\beta}M_{\alpha\beta}+C^{\alpha\beta}\bar{M}_{\alpha\beta}X \) \;.
}
Thus, one can use $\L^{S^{2}}_{\Bbb X}$ itself for localization, which is an important simplification in evaluating the one-loop determinant using spherical harmonics.%
\footnote{Strictly speaking, one should use a $\Q_A$-exact action which is positive definite, so that one localizes to the zero-locus. $\L^{S^2}_{\Bbb X}$ has positive definite real part, provided that $0 \leq q \leq 2$.}

\subsection*{\it Same Representation}

Let us consider first the case of a pair of semichiral fields in representation $(\fR, \fR)$ with the flat-space action (\ref{flat space lagrangian left and right}). Since so far we treated $\bar{\Bbb X}$ and $\Bbb X$ as independent fields, we can use the result (\ref{action general sphere}) for each individual term in (\ref{flat space lagrangian left and right}) and the Lagrangian on $S^2$ is therefore given by
\bea
\label{general lagrangian}
\L_{LR}^{S^{2}} &= D_{\mu}\bar{X}^{i}D^{\mu}X_{i}+\bar{X}^{i}\(\sigma_1^{2}+\sigma_2^{2} + i D \)X_{i} +\bar{F}^{i}F_{i} \\
&\quad -\bar{M}_{\alpha\beta}^{i}M^{\beta\alpha}_{i} - \bar{X}^{i}\P_{\alpha\beta}M^{\alpha\beta}_{i} + \bar{M}^{\alpha\beta,i}\P_{\beta\alpha}X_{i} \\
&\quad -i\bar{\psi}^{i}\gamma^{\mu}D_{\mu}\psi_{i} + \bar{\psi}^{i}\(i\sigma_1 - \gamma_{3}\sigma_2\)\psi_{i} + i \bar{\psi}^{i}\lambda X_{i} - i \bar{X}^{i}\bar{\lambda}\psi_{i} - \eta^{i}\psi_{i} - \bar{\psi}^{i}\bar{\eta}_{i} \\ 
&\quad +i\bar{\chi}_{i}\gamma^{\mu}D_{\mu}\chi^{i} - \bar{\chi}_{i}\(i\sigma_1 - \gamma_{3}\sigma_2\)\chi^{i} + i \bar{X}^{i}\lambda\bar{\chi}_{i} - i\chi^{i}\bar{\lambda}X_{i} \\
&\quad +\tfrac{iq}{r}\bar{X}^{i}\sigma_1 X_{i} +\tfrac{q(2-q)}{4r^2}\bar{X}^{i}X_{i} -\tfrac{q}{2r} \Big( \bar{\psi}^{i}\psi_{i} + \chi^i\bar{\chi}_i + \bar{X}^{i}C^{\alpha\beta}M_{i\, \alpha\beta}+C^{\alpha\beta}\bar{M}_{\alpha\beta}^{i}X_{i} \Big) \;,
\eea
where the flavor indices $i=(L,R)$ are contracted with
\equ{
\M_{\bar\imath  j} = - \pmat{ 1 & \alpha \\ \alpha & 1 } \;.
} 
The action (\ref{general lagrangian}) is $\Q_A$-exact, being a sum of $\Q_A$-exact terms.

In the simple model with a single pair of semichiral fields and gauge group $U(1)$, the R-charge $q$ is unphysical, because it can be set to the canonical value $q=0$ by mixing the R-current with the gauge current $\big($this is no longer true if we have multiple semichiral pairs charged under the same $U(1)\big)$. However, we keep $q$ for now and set it to zero only at the end of the calculation; this will reduce the number of BPS configurations to be taken into account in the localization.

\subsection*{\it Conjugate Representations}

Let us move to the case of a pair of semichiral fields in conjugate representations, whose flat-space Lagrangian includes off-diagonal terms of the form $\Bbb X_{L} \Bbb X_{R}$ appearing in (\ref{flat space lagrangian left and right opposite charges}):
\equ{\label{components action opposite charges}
\L_\beta^{\Bbb R^{2}} = \beta \int d^{4}\theta\, \big( \Bbb X_{L} \Bbb X_{R} + \bar{\Bbb X}_{L} \bar{\Bbb X}_{R} \big) = \beta \( M^{\alpha\beta} M_{\alpha\beta} - \bar\eta \bar\chi + \bar{M}^{\alpha\beta} \bar{M}_{\alpha\beta} - \eta\chi \).
}
This Lagrangian is actually invariant under the $S^{2}$ SUSY transformations (\ref{SUSY variations sphere 1}), with no need for $\frac1r$ improving terms, \ie{}, $\L_\beta^{S^2} = \L_\beta^{\Bbb R^2}$. Furthermore, it is also $\Q_A$-exact, namely:
\eqs{\bep \e \int d^{2}x \, \mathcal L_{\beta}^{S^{2}} = \delta_{\e}\delta_\bep \int d^{2}x \, \Big(& X_{L} M_{+-}-X_{R}M_{-+} +\bar \chi_{-}\psi_{+}^{R}+\psi_{-}^{L}\bar \chi_{+} - \tfrac{1}{r} X_{R} X_{L} \nonumber\\ 
&+ \bar X_{L} \bar M_{+-}-\bar X_{R} \bar M_{-+} +\chi_{-}\bar \psi_{+}^{R}+\bar \psi_{-}^{L}\chi_{+} - \tfrac{1}{r} \bar{X}_{L}\bar X_{R}\Big)\,.
\label{exact action opposite charges}
}
Note that although $\frac{1}{r}$ terms appear inside the integral on the right-hand side, these are cancelled against $\frac{1}{r}$ terms coming from $\delta_{\e} \delta_\bep$. Summarizing, the generalization of (\ref{flat space lagrangian left and right opposite charges}) to $S^2$ is the sum of (\ref{general lagrangian}) (where flavor indices are contracted instead with $\M_{\bar\imath j}=\delta_{\bar\imath j}$) and (\ref{components action opposite charges}).

Since a general Lagrangian for a number $N_F$ of semichiral pairs can always be rewritten as $N_F$ decouped copies of a single pair, the elements given above are sufficient to write the general action on $S^2$ for any number of semichiral multiplets.

\section{Localization on the Coulomb Branch}\label{Localization on the Coulomb Branch}

In this section, we compute the $S^2$ partition function of the gauge theories at hand by means of localization on the Coulomb-branch localization. 

We wish to compute the path integral
\equn{
Z_{S^2}=\int \mathcal D\varphi  \, e^{-\S[\varphi]} \;,
}
where $\varphi$ are all fields in the theory, namely those in vector multiplets, semichiral multiplets, and possibly chiral multiplets. The action is given by
\equ{
\S=\int d^2x \big( \L_\text{VM} +\L_\text{FI} + \L_\text{chiral} + \L_\text{semichiral} \big) \;,
}
where each term is the appropriate Lagrangian on $S^2$ and $\L_\text{FI}=- i \xi D+\frac{i\theta}{2 \pi}F_{12}$ is the standard FI term (which needs no curvature couplings on $S^2$). To perform the localization, we pick a Killing spinor $\epsilon$ (our choice is in (\ref{expression explicit killing spinor}), but all choices are equivalent up to rotations of $S^2$) and call $\Q_A$ the generator of supersymmetry variations along $\epsilon$ and $\bar\epsilon = \epsilon^c$, as in \cite{Benini:2012ui}. The supercharge $\Q_A$ generates an $SU(1|1)$ superalgebra, that we will use for localization. Following the usual arguments \cite{Witten:1988xj, Witten:1991zz}, the partition function localizes on the BPS configurations given by $\{ \Q_A \cdot \text{fermions}=0 \}$, and it is given exactly by the one-loop determinant around such configurations. The contribution from vector and chiral multiplets was studied in \cite{Benini:2012ui, Doroud:2012xw}, that from twisted chiral multiplets in \cite{Gomis:2012wy}, and that from twisted vector multiplets in \cite{Doroud:2013pka}. Here we study the contribution from semichiral multiplets.

We begin by studying the BPS equations.  These follow from setting the variations of all fermions in (\ref{SUSY variations sphere 1}) to zero and are analyzed in detail in Appendix~\ref{BPS equations for the semichiral multiplet}. We show that for a generic value of $q\neq0$, the only smooth solution  is
\equ{
\label{BPS Chiral}
X= \bar X = F = \bar F = M_{\alpha\beta} = \bar M_{\alpha\beta} = 0 \;.
}
Thus, like in the case of chiral multiplets, the BPS configuration for generic $q$ is only the trivial one. The BPS configurations for the vector multiplet are given by \cite{Benini:2012ui, Doroud:2012xw} 
\equ{
\label{background vector}
0 = F_{12} - \frac{\sigma_2}{r} = D + \frac{\sigma_1}{r} = D_{\mu}\sigma_1 = D_\mu \sigma_2 = [\sigma_1,\sigma_2] \;.
}
Flux quantization of $F_{12}$ implies that, up to gauge transformations, $\sigma_2 = \frac{\fm}{2r}$ where $\fm$ is a co-weight (\ie{} $\fm$ belongs to the Cartan subalgebra of the gauge group algebra and $\rho(\fm) \in \bZ$ for any weight $\rho$ of any representation $\fR$). Thus, the set of BPS configurations are parametrized by the continuous variable $\sigma_1$ and the discrete fluxes $\fm$.

As shown in the previous section, the kinetic actions for semichiral fields are $\Q_A$-exact. Thus, we can use the kinetic actions themselves as a deformation term for localization and we should compute the one-loop determinants arising from those actions. We now compute the determinant for semichiral fields in the same gauge representation, as well as in conjugate representations; the determinants coincide, as they should, since such theories are related by a simple change of variables.

\subsection{One-loop Determinants}\label{One-loop Determinants}

\subsection*{\it Same Representation}

Consider the Lagrangian (\ref{general lagrangian}) and expand it at quadratic order around the BPS background. Let us look at bosonic fields first, using the basis
\eqsn{
\X = \Big( X^L,X^R,M^L_{-+},M^L_{--},M^R_{++},M^R_{+-}, F^L, F^R \Big)^\sT \;.
}
The bosonic part of the quadratic action is given by $\bar{\X} \O_{B} \X$, where $\O_{B}$ is the $8\times 8$ operator
\equ{\label{OXMF}
\O_{B} = \pmat{
\O_X & \alpha\,\O_X & \tfrac{q}{2r} -{\s} & 2 i D_{++} & -\alpha\, 2 i D_{--} & -\alpha\(\tfrac{q}{2r} + \bar{{\s}}\) & 0 & 0\\
\alpha\,\O_X & \O_X & \alpha\(\tfrac{q}{2r} -{\s}\) & \alpha\, 2 i D_{++} & -2 i D_{--} &  -\(\tfrac{q}{2r} +\bar{{\s}}\) & 0 & 0\\
\alpha\( \sigma -\tfrac{q}{2r} \) & -\tfrac{q}{2r} + {\s} & 0 & 0 & 0& -1 & 0 & 0\\
\alpha\, 2 i D_{--} & 2 i D_{--} & 0 & \alpha & 0& 0 & 0 & 0\\
-2 i D_{++} &  - \alpha \, 2 i D_{++} & 0 & 0 & \alpha & 0 & 0 & 0\\
\tfrac{q}{2r} +\bar{{\s}} & \alpha \(\tfrac{q}{2r} +\bar{{\s}}\)& -1 & 0 & 0& 0 & 0 & 0\\
0 & 0 & 0 & 0 & 0 & 0 & 1 & \alpha \\
0 & 0 & 0 & 0 & 0 & 0 & \alpha & 1}
}
and
\equ{\label{operator O_X}
\O_X = -\Box + \sigma_1^2 + \sigma_2^2 + i \frac{(q-1)\sigma_1}r + \frac{q (2-q)}{4r^2} \;.
}
All appearances of $\sigma_1, \sigma_2$ in this matrix (and all matrices below) are to be understood as $\rho(\sigma_1)$, $\rho(\sigma_2)$, but we omit this to avoid cluttering the matrices; we will reinstate them in the expressions for the determinants below. The analysis of the eigenvalues of (\ref{OXMF}) contains different cases, depending on the angular momentum $j$ on $S^2$. Assuming $\alpha\neq 0$, and putting all cases together, the determinant in the bosonic sector is given by (see Appendix~\ref{1-loop Determinants}):
\begin{multline}
\text{Det}\,\O_B = \prod_{\rho\in \fR} \frac{ \alpha^{|\rho(\fm)|-1} }{ \alpha^{|\rho(\fm)|+1} } \prod_{j=\frac{|\rho(\fm)|}{2}}^\infty \bigg[j^2 + \frac{(\alpha^2 - 1)\rho(\fm)^2}4 -\alpha^2 \Big( \frac q2 -i r \rho(\sigma_1) \Big)^2 \bigg]^{2j+1} \times \\
\times \bigg[(j+1)^2 + \frac{(\alpha^2 - 1) \rho(\fm)^2}4 - \alpha^2 \Big( \frac q2 -i r \rho(\sigma_1) \Big)^2 \bigg]^{2j+1} \bigg( \frac{(\alpha^2 -1)^2}{r^4} \bigg)^{2j+1} \;.
\end{multline}
Now we turn to the fermionic determinant. In the basis
\equn{
\Psi = \Big( \psi^{L+},\psi^{L-},\psi^{R+},\psi^{R-},\bar{\eta}^{L+},\bar{\eta}^{R-},\bar{\chi}^{L+},\bar{\chi}^{R-} \Big)^\sT \;,
}
the  quadratic action for fermions is  $\bar{\Psi}\O_{F}\Psi$ where $\O_F$ reads
\equn{\pmat{
-\(\tfrac{q}{2r}+\bar{{\s}}\) & 2 i D_{--} & -\alpha \big( \tfrac{q}{2r} + \bar{ {\s}} \big) & 2 i \alpha D_{--} & -1 & 0 & 0 & 0\\
-2 i D_{++} & \tfrac{q}{2r}-{\s} & -2 i \alpha D_{++} & \alpha\(\tfrac{q}{2r}-{\s}\) & 0 & \alpha & 0 & 0\\
-\alpha \big( \tfrac{q}{2r}+\bar{{\s}} \big) & 2 i \alpha D_{--} & -\(\tfrac{q}{2r}+\bar{{\s}}\) & 2 i D_{--} & -\alpha & 0 & 0 & 0\\
-2 i \alpha D_{++} & \alpha \big( \tfrac{q}{2r}-{\s} \big) & -2 i D_{++} & \tfrac{q}{2r}-{\s} & 0 & 1 & 0 & 0\\
-\alpha & 0 & -1 & 0 & 0 & 0 & 0 & 0\\
0 & 1 & 0 & \alpha & 0 & 0 & 0 & 0\\
0 & 0 & 0 & 0 & 0 & 0 & -\alpha \big( \tfrac{q}{2r} - {\s} \big) & -2 i D_{--}\\
0 & 0 & 0 & 0 & 0 & 0 & 2 i D_{++} & \alpha \big( \frac{q}{2r} + \bar{{\s}} \big)}.
}
An analysis of the eigenvalues of this operator gives the determinant
\begin{multline}
\text{Det}\,\O_F = \prod_{\rho\in \fR} \bigg( \frac{ (1 - \alpha^2) \alpha^2}{r^2} \bigg)^{|\rho(\fm)|} \bigg[ \frac{\rho(\fm)^2}4 - \Big( \frac{q}{2}-i r \rho(\s_1) \Big)^2 \bigg]^{|\rho(\fm)|} \times \\
\times \prod_{j=\frac{|\rho(\fm)| + 1}2}^\infty \bigg( \frac{\alpha^2 - 1}{r^2} \bigg)^{4j+2} \bigg[ \left( j+\frac{1}{2} \right)^2 + \frac{(\alpha^2 - 1) \rho(\fm)^2}4 -\alpha^2 \Big( \frac{q}{2} -i r \rho(\sigma_1) \Big)^2 \bigg]^{4j+2} \;.
\end{multline}
Bringing the bosonic and fermionic determinants together leads to many cancellations and the final result is%
\equ{\label{semichiralPF}
Z_{LR} =\frac{\text{Det}\,\O_{F}}{\text{Det}\,\O_{B}}= \prod_{\rho \in \fR}  \frac{(-1)^{|\rho(\fm)|}}{\frac{\rho(\fm)^2}{4}-\(\frac{q}{2}-i r \rho(\s_1)\)^2} \;.
}
Note that in this expression, the dependence on the parameter $\alpha$ has cancelled, as  expected from the fact that this parameter appears in a $\Q_A$-exact term.\footnote{In fact, here we have dropped overall factors of $r^2$ and $(\alpha^{2}-1)$, which can be reabsorbed in the path-integral measure.}

This expression has simple poles at $\rho(\sigma_1)= -\frac{i}{2r}\(q\pm \rho(\fm)\)$ for  $\rho(\fm)\neq 0$, and a double pole at $\rho(\sigma_1)=-\frac{i q}{2r}$ for $\rho(\fm)=0$. Using properties of the $\Gamma$-function, (\ref{semichiralPF}) can be written as
\eqs{\label{semichiralPF same charges}
Z_{LR}= \prod_{\rho \in \fR} \frac{\Gamma \left(\frac{q}{2} - i r \rho(\sigma_1) - \frac{\rho(\fm)}{2} \right)}{\Gamma \left(1 - \frac{q}{2} + i r \rho(\sigma_1) - \frac{\rho(\fm)}{2} \right)} \cdot \frac{\Gamma \left(-\frac{q}{2} + i r \rho(\sigma_1) + \frac{\rho(\fm)}{2} \right)}{\Gamma \left(1 + \frac{q}{2} - i r \rho(\sigma_1) + \frac{\rho(\fm)}{2} \right)} \;.
}
The radius $r$ can be reabsorbed into $\sigma_1$, making it into a dimensionless variable. In fact (\ref{semichiralPF same charges}) coincides with the one-loop determinant for two chiral fields in conjugate representations of the gauge group, opposite R-charges, and no twisted mass parameter turned on. Each $\Gamma$-function in the numerator has an infinite tower of poles, most of which cancel against the poles of the denominator (such a cancellation does not occur for a pair of chiral multiplets with generic twisted masses and R-charges).

\subsection*{\it Conjugate Representations}

Here we compute the one-loop determinant for semichiral fields in representation $(\fR, \wb\fR)$. The flat-space Lagrangian is the semichiral dual to the one in the previous section  (see Appendix \ref{Semichiral-Semichiral Duality}). To place it on $S^{2}$ we take (\ref{general lagrangian}), where the indices are contracted appropriately, and the off-diagonal term (\ref{components action opposite charges}). To second order in the fluctuations, the bosonic action is $\bar{\X}\wt \O_B\X$ where
\eqsn{
\X&=\left(X^L, \bar{X}^R, M^L_{-+}, M^L_{--}, \bar{M}^R_{++}, \bar{M}^R_{+-},F^L, \bar{F}^R\right)^\sT
}
and
$$
\wt\O_B = \pmat{
\alpha_s \O_X^{(q)} & 0 & \alpha_s \(\tfrac{q}{2r} -{\s}\) & \alpha_s 2 i D_{++} & 0 & 0 & 0 & 0\\
0 & \O_X^{(-q)} & 0 & 0 & -2 i D_{--} & \tfrac{q}{2r} -{\s} & 0 & 0\\
0 & \tfrac{q}{2r} +\bar{{\s}} & -\alpha & 0 & 0& -1 & 0 & 0\\
0 & 2 i D_{--} & 0 & \alpha & 0& 0 & 0 & 0\\
- \alpha_s 2 i D_{++} & 0 & 0 & 0 & \alpha & 0 & 0 & 0\\
\alpha_s \(\tfrac{q}{2r}+\bar{{\s}}\) & 0 & -\alpha_s & 0 & 0 & -\alpha & 0 & 0\\
0 & 0 & 0 & 0 & 0 & 0 & \alpha_s & 0\\
0 & 0 & 0 & 0 & 0 & 0 & 0 & 1} \;.
$$
in which we defined $\alpha_s = \alpha^2-1$ and $\O_X^{(q)}$ is the operator in (\ref{operator O_X}) using R-charge $q$. The second-order fermionic action is $\bar{\Psi}\wt\O_F \Psi$ with
\eqsn{
\Psi&= \left( \psi^{L+}, \psi^{L-}, \bar{\psi}^{R+}, \bar{\psi}^{R-}, \bar{\eta}^{L+}, \eta^{R-}, \bar{\chi}^{L+}, \chi^{R-} \right)^\sT
}
and $\wt\O_F$ equals
$$
\pmat{
- \alpha_s \(\tfrac{q}{2r}+\bar{{\s}}\) & \alpha_s 2 i D_{--} & 0 & 0 & -\alpha_s & 0 & 0 & 0\\
-\alpha_s 2 i D_{++} & \alpha_s \(\tfrac{q}{2r}-{\s}\) & 0 & 0 & 0 & 0 & 0 & 0\\
0 & 0 & \tfrac{q}{2r}-{\s} & 2 i D_{--} & 0 & 0 & 0 & 0\\
0 & 0 & -2 i D_{++} & -\(\tfrac{q}{2r}+\bar{{\s}}\) & 0 & -1 & 0 & 0\\
0 & 0 & 1 & 0 & 0 & 0 & \alpha & 0\\
0 & \alpha_s & 0 & 0 & 0 & 0 & 0 & -\alpha\\
0 & 0 & 0 & 0 & \alpha & 0 & 0 & -2 i D_{--}\\
0 & 0 & 0 & 0 & 0 & -\alpha & \alpha_s 2 i D_{++} & 0} \;.
$$
Evaluating the determinants, we again find \eqref{semichiralPF same charges}, as expected.

\subsection{Integration Contour and Instanton Corrections}\label{Integration Contour}

The full partition function requires integration over the zero-mode $\sigma_1$, as well as summation over the flux sectors $\fm$ (which are co-weights of the gauge group):
\equ{\label{full partition integral}
Z_{LR}^{S^{2}} = \frac{1}{|\mathcal W|} \sum_{\fm} \int \frac{d \sigma_1}{2 \pi} \; e^{- 4 \pi i \xi \Tr \sigma_1 - i \theta \Tr \fm} \; Z_\text{gauge} \;  Z_{LR} \;,
}
where $Z_{LR}$ is given in (\ref{semichiralPF same charges}), $|\mathcal W|$ is the order of the Weyl group, and  $ Z_{\text{gauge}}$ is the contribution from the vector multiplet given in \cite{Benini:2012ui, Doroud:2012xw}. The integral is over some contour in the complex plane which needs to be specified, as we are going to discuss.

Let us make our point through a concrete example. Consider $N_{F}$ pairs of semichiral fields coupled to a $U(1)$ gauge field, each pair having charges $(1,-1)$ and R-charges $(q,-q)$. From now on we set the radius $r=1$ to avoid cluttering the formul\ae. We can simply take the result (\ref{semichiralPF same charges}) for each pair and obtain
\eqs{\nonumber
Z_{LR}^{S^{2}}(\xi,\theta) = \sum_\fm \int \frac{d \sigma_1}{2 \pi}  e^{- 4 \pi i \xi \sigma_1 - i \theta \fm} \, \bigg(\frac{\Gamma \left(\frac{q}{2} - i  \sigma_1 - \frac\fm2 \right)}{\Gamma \left(1 - \frac{q}{2} + i  \sigma_1 - \frac\fm2 \right)} \cdot \frac{\Gamma \left(-\frac{q}{2} + i  \sigma_1 + \frac\fm2 \right)}{\Gamma \left(1 + \frac{q}{2} - i  \sigma_1 + \frac\fm2 \right)} \bigg)^{N_{F}} \;.
}
We  have not included any twisted mass.
As we have discussed, the integrand coincides with that of $N_{F}$ chiral fields of charge $1$ and $N_{F}$ chiral fields of charge $-1$. It has poles of order $N_{F}$ at $\sigma_1= -\frac{i}{2} ( q\pm \fm )$ for  $\fm\neq 0$, and a pole of order $2N_{F}$ at $\sigma_1= - \frac i2 q$ for $\fm=0$. If the R-charge is set to $q=0$, in the $\fm=0$ sector one encounters a pole on the real line. Unless this pole is avoided by the integration contour, it leads to a divergence; such a divergence is physical, as it is a manifestation of the non-compactness of the target space. To avoid the divergence and extract some useful information, the contour prescription should be modified.

The way this is dealt with in the case of chiral fields is to introduce twisted masses. For instance, in the case $N_F=1$ of two chiral fields $\Phi$, $\tilde \Phi$ there is a $U(1)_g \times U(1)_F$ symmetry: the first $U(1)_g$ is gauged, while the second remains as a global flavor symmetry. One can turn on a twisted mass $\wt M$ associated to $U(1)_F$ which splits the double pole for $\fm=0$ into two poles, one above the real axis and one below it. Integrating along the real line (\ie, going through the split poles for $\fm=0$) leads to a finite result, regulated by $\wt M$. From the point of view of the low-energy NLSM, a twisted mass corresponds to a quadratic potential on the target which effectively compactifies the model. 

Unfortunately, in the semichiral case this resolution is not possible. As we have discussed, the only $U(1)$ symmetry in these models is the one being gauged%
\footnote{As discussed in \cite{Crichigno:2015pma}, there is a way to introduce additional mass parameters in these theories using the SVM. This is achieved by giving the chiral field strength (of R-charge 2) in the SVM a VEV, but since this field has a non-zero R-charge, this breaks A-type supersymmetry on the sphere.}%
---the would-be flavour symmetry $U(1)_F$ is broken by the off-diagonal terms in (\ref{flat space lagrangian left and right}) or (\ref{flat space lagrangian left and right opposite charges}). Thus, we are forced to choose a different contour prescription to avoid the singularity. One possible choice is to simply go around the double pole for $\fm=0$, either above or below it, avoiding the classical (divergent) contribution; this is the contour prescription we choose.

From the point of view of the large-volume NLSM, at least in the Abelian case, the prescription can be interpreted as follows. Going back to the model with chiral fields, suppose that $\xi<0$ which selects a particular large-volume limit. One can rewrite the integral along the real line as a sum of the residues at the poles in the upper half-plane. Each residue can be interpreted as the contribution from a different instanton sector to the NLSM path-integral \cite{Jockers:2012dk, Honma:2013hma, Bonelli:2013mma}, and in particular the residue at the pole collapsing with its partner (as $\wt M \to 0$) reproduces the classical and one-loop contribution, divergent in the limit. We choose, instead, a contour that goes above the collapsing pair of poles: this essentially misses the collapsing pole, and therefore it computes the NLSM path-integral without the zero-instanton contribution. This makes sense because the space of NLSM field configurations is the disconnected sum of instanton sectors, therefore it is a well-defined operation to remove one of them from the path-integral. Had we considered the large-volume limit at $\xi>0$, we would have taken the contour that goes below the collapsing pair of poles. Notice that the instanton corrections captured by this prescription are identical to the ones in the corresponding K\"ahler case.

Performing the contour integral in the simple Abelian model above, with $\xi < 0$, using Cauchy's integral formula and summing over $\fm$, one finds
\begin{multline}
Z^\text{inst}_{N_F} (\xi,\theta) = \sum_{j=0}^{N_F-1}\frac{(2N_F-2-j)!}{(N_F-1)!^2}\binom{N_F-1}{j}\(-4\pi\xi\)^j\times \\
\times\left[\text{Li}_{2N_F-1-j}\((-1)^{N_F}e^{i \theta+2\pi\xi}\)+\text{Li}_{2N_F-1-j}\((-1)^{N_F}e^{-i \theta+2\pi\xi}\) \right] \;,
\end{multline}
where we have set $q=0$ after performing the integral.

A case of particular interest is $N_F=2$ $\big($or $N_F=1$, and in addition two chiral fields with gauge charges $(1,-1)\big)$, for which the target space has complex dimension three.  As we discuss in Section~\ref{Example: Generalized Conifold}, this gauge theory describes a conifold with a generalized K\"{a}hler structure. The contribution due to instantons reads
\equ{
\label{instanton corrections conifold}
Z^\text{inst}_\text{conifold} (\xi,\theta)=2\(\text{Li}_{3} (e^{2 \pi  \xi +i \theta })+\text{Li}_{3} (e^{2 \pi  \xi -i \theta })\)-4 \pi \xi\(\text{Li}_{2}(e^{2 \pi  \xi +i \theta })+\text{Li}_{2}(e^{2 \pi  \xi -i \theta }) \) \;.
}
We discuss instanton corrections from the point of view of the NLSM in Section~\ref{Sigma Models on Generalized Kahler Manifolds}, but before that we make a comment on gauge theories with other vector multiplets.

\subsection{Coupling to Other Vector Multiplets}\label{Coupling to Other Vector Multiplets}

As mentioned earlier, a salient feature of semichiral fields is that they can couple minimally to various vector multiplets. So far we have studied the coupling to the usual vector multiplet, but they may also couple to the twisted  vector and to the SVM. For the SVM one can define two gauge-invariant field strengths $(\Bbb F, \tilde{\Bbb F})$, which are chiral and twisted chiral, respectively (see Appendix~\ref{N=(2,2) Multiplets} for a brief overview). In the Abelian case, a gauge-invariant action in flat space is:
\begin{multline}
\label{SVM gauged action opposite charge Nf unconstrained}
\L_\text{SVM, $LR$} = - \frac{1}{2e^{2}}\int d^{4}\theta\, \Big( \bar {\tilde{ \Bbb F}} \tilde{\Bbb F} -\bar {\Bbb F} \Bbb F \Big)  +\(i \int d^{2}\theta \,  s\, \Bbb{ F}+c.c.\)+\(i \int d^{2}\tilde \theta \,  t\, \Bbb{\tilde F}+c.c.\)\\
+\sum_{i=1}^{N_{F}}\int d^{4}\theta
\left[ \Bbb{ \bar X}_L^{i} e^{Q_{i}V_L } \Bbb{ X}_L^{i} + \Bbb{ \bar X}_R^{i} e^{Q_{i}V_R} \Bbb{ X}_R^{i}+ \beta_{i} \big( \Bbb{  X}_L^{i} e^{ -iQ_{i}{\Bbb V}} \Bbb{ X}_R^{i}+c.c. \big)  \right]  \;.
\end{multline}
The first line is the kinetic action for  the SVM and $s, t$ are two complex FI parameters. The Higgs branch has complex dimension $2 N_{F}-2$  and for generic value of the parameters $\beta_{i}$ the geometry is generalized K\"{a}hler  \cite{Crichigno:2015pma}. For a special choice of the parameters the geometry becomes hyperk\"{a}hler \cite{Crichigno:2011aa}. 

Now, we can promote $s$ in (\ref{SVM gauged action opposite charge Nf unconstrained}) to a chiral field $\Phi$:
$$
s \quad \rightarrow \quad \Phi \;.
$$
As discussed in \cite{Crichigno:2015pma}, by doing so $\Phi$ acts as a Lagrange multiplier imposing  $\Bbb F=0$ and by choosing an appropriate SVM gauge, the action reduces to (\ref{flat space lagrangian 1,-1 charges}), namely the action of semichiral fields coupled to the usual vector multiplet. From the superspace point of view, this is a more natural way to formulate the gauge theories (\ref{flat space lagrangian 1,-1 charges}). For our purposes here, however, it is simpler to derive the couplings to curvature when working in the formulation (\ref{flat space lagrangian 1,-1 charges}).
Note that the theory in (\ref{SVM gauged action opposite charge Nf unconstrained}) and the one with dynamical $\Phi$ only differ by a superpotential term, moreover the Lagrange multiplier must be neutral and with R-charge $0$. It follows that the partition functions in $\Q_A$-localization for the two Abelian theories coincide.

\section{Sigma Models on Generalized Kähler Manifolds}\label{Sigma Models on Generalized Kahler Manifolds}

The GLSMs discussed in this paper realize NLSMs on generalized K\"{a}hler manifolds. In this section we discuss instanton corrections from the point of view of the NLSM on the generalized K\"ahler space, and comment on possible interpretations for the GLSM partition function. We first review some known results on topologically twisted sigma models with K\"{a}hler and generalized K\"{a}hler target spaces, and then discuss some interesting issues that the localization computation raises for the models at hand.

\subsection{Topologically Twisted Theories}\label{Topological theories}

The topological twist is a powerful method to obtain exact results about specific sectors in supersymmetric theories \cite{Witten:1988xj}. After a topological twist, a linear combination of the supercharges becomes a scalar charge $\mathcal Q_\text{BRST}$, referred to as the BRST operator. In the case of nonlinear sigma models with a K\"{a}hler target space, the classical action of the NLSM can be written as the sum of a BRST-exact term and a topological term \cite{Witten:1991zz}. The path-integral localizes on fixed points of $\Q_\text{BRST}$ and therefore the partition function is given by a sum over such configurations, with a weight given by the exponential of the topological term. In the A-model, the fixed points of $\mathcal Q_\text{BRST}$ are holomorphic maps, while in the B-model they are constant maps. 

The topologically twisted versions of sigma models with a generalized K\"{a}hler target space were first studied in \cite{Kapustin:2004gv} (see \cite{Bredthauer:2006hf, Chuang:2006vt, Zucchini:2006ii} for further discussions). A BRST operator can be constructed and its fixed points are given (in the case of an A-twist) by configurations satisfying
\equ{
\label{generalized holomorphic equations}
\tfrac{1}{2} \big( 1- i J_{+} \big) \, \bar \partial X=0 \;,\qquad\quad \tfrac{1}{2} \big( 1+i J_{-} \big) \, \partial X=0 \;,
}
where $X$ are \textit{real} coordinates on the target space and target space indices have been omitted. These equations are a generalization of both the usual holomorphic maps of the ordinary A-model and the constant maps of the ordinary B-model. 

As in the K\"{a}hler case, it is natural to expect that the classical action can be written as a sum of a BRST-exact term and a topological term. This was proven when $[J_{+},J_{-}]=0$, and conjectured to also hold when $[J_{+},J_{-}]\neq0$ \cite{Hull:2008vs}.

Here we wish to focus on the space of solutions to (\ref{generalized holomorphic equations}). As discussed in \cite{Kapustin:2004gv}, for generic $J_{\pm}$, and at a generic point on the manifold, these equations are more restrictive than the ordinary holomorphic map condition: in fact the only solutions are constant maps and non-trivial instanton solutions are not possible. However, if $J_+ = J_-$ then the two equations in (\ref{generalized holomorphic equations}) become complex conjugates to each other and they reduce to the standard holomorphic map condition, allowing for non-trivial instanton solutions. Thus, instanton corrections can only arise from (compact) submanifolds on which the pull-back of $\omega ≡ g (J_+ - J_-)$ is degenerate.\footnote{Similarly, in the case of a B-twist instanton corrections arise only from compact submanifolds on which $g(J_+ + J_- )$ is degenerate.}

We now discuss in more detail the generalized K\"{a}hler structure realized by the GLSMs discussed in a particular example.

\subsection{Example: A Generalized Kähler Metric on the Conifold}\label{Example: Generalized Conifold}

Consider a pair of semichiral fields and a pair of chiral fields $(\Bbb X_{L}, \Bbb X_{R},\Phi_{1}, \Phi_{2})$,  charged under a $U(1)$ vector multiplet with charges $(1,-1,1,-1)$, respectively.%
\footnote{One may consider a GLSM involving semichiral fields only. This model is discussed in more detail in \cite{Crichigno:2015pma} but requires the introduction of a new vector multiplet and calculations are less straightforward. Nonetheless, the discussion below also applies to those models.}
The kinetic Lagrangian is given by
\equ{
\label{gauge action generalized conifold}
\L = \L_\text{VM} + \int \! d^{4}\theta \Big( \bar{\Bbb X}_{L} \Bbb X_{L}+\bar{\Bbb X}_{R} \Bbb X_{R} + \beta \big( \Bbb X_{L} \Bbb X_{R} + \bar{\Bbb X}_{L} \bar{\Bbb X}_{R} \big) + \bar \Phi_{1}  \Phi_{1}+\bar \Phi_{2}\Phi_{2} \Big) + \frac{i}{2} \int \! d^2\tilde \theta \; t \, \Sigma + c.c. \;,
}
where $\L_\text{VM}$ is given in (\ref{action vector multiplet}), $t= i \xi+\tfrac{\theta}{2\pi}$ and we take $\xi>0$. As discussed in Section~\ref{NLSM description} (and including the usual contribution from chiral fields), the space of vacua of this theory is given by solutions to 
\equ{
\label{moment map generalized conifold}
|X_{L}|^{2}+|\phi_{1}|^{2}- |X_{R}|^{2}-|\phi_{2}|^{2}= \xi \;,
}
modulo $U(1)$ gauge transformations. As a quotient space, this is the description of the resolved conifold, a $\Bbb C^{2}$ bundle over $\Bbb{CP}^{1}$ \cite{Candelas:1989js}. For $\xi=0$ the space has a conical singularity at the origin (the tip of the cone), but for finite $\xi$ the singularity is blown-up to an $S^{2}$ of radius $|\xi|^{\frac{1}{2}}$. For $\xi>0$, the $S^{2}$ at the tip is given by setting $X_{R}=\phi_{2}=0$. Gauge-invariant combinations are 
\equ{
\label{gauge inv coords}
\Bbb X_L'=\frac{\Bbb X_L}{\Phi_1} \;,\qquad\qquad \Bbb X_R'=\Phi_1 \Bbb X_R \;,\qquad\qquad \Phi= \Phi_1 \Phi_2 \;.
}
These are good coordinates on the target in the patch $\Phi_1\neq 0$.

It is well known that this space admits a K\"{a}hler metric (in fact, a Calabi-Yau metric) and can be realized by a GLSM with four chiral fields with charges $(1,1,-1,-1)$ \cite{Klebanov:1998hh}. In the description (\ref{gauge action generalized conifold}) this corresponds to the limit $\beta \rightarrow \infty$.

The genus-zero partition function of the topological string on the conifold was computed in \cite{Candelas:1990rm}.%
\footnote{For a review of the results for higher genus contributions and the interesting relation to 3d Chern-Simons theory, see \cite{Gopakumar:1998ki, Ooguri:1999bv} and references therein.}
In addition to the classical contribution it contains nontrivial instanton corrections, which arise from multi-coverings of the worldsheet onto the $S^{2}$ at the tip. In supersymmetric localization these are captured by (\ref{instanton corrections conifold}). 

As mentioned in the introduction, the full geometric data is encoded in the NLSM's $\N=(2,2)$ superspace Lagrangian $K$---the generalized  K\"ahler potential. This function can be easily computed in the UV by classically integrating out the vector multiplet $V$. In the limit $e\rightarrow \infty$, the equation of motion for the vector multiplet is simply a quadratic equation for $e^V$. Solving it and plugging $e^V$ back into the action (\ref{gauge action generalized conifold}), one obtains the generalized K\"{a}hler potential
\equ{
\label{generalized potential conifold}
K_\text{UV} = \sqrt{\xi^2+4\,r^{2}} - \xi \, \log \bigg( \frac{\xi+\sqrt{\xi^2+4\,r^{2}}}{1+|\Bbb X_{L}|^{2}} \bigg) + \beta \big( \Bbb X_{L} \Bbb X_{R} + \bar{\Bbb X}_{L} \bar{\Bbb X}_{R} \big) \;,
}
where we have written the potential in terms of the gauge-invariant coordinates (\ref{gauge inv coords}) (and dropped the primes) and defined the radial coordinate $
r^{2}\equiv \big( 1+|\Bbb X_{L}|^{2} \big) \big( |\Bbb X_{R}|^{2}+|\Phi|^{2} \big)$. Both square roots  in (\ref{generalized potential conifold}) should be taken as the positive root, since we have chosen a particular branch of the solution corresponding to a real $e^V$. The subscript UV reminds us that, although this is a description of the gauge theory at low energies with respect to the gauge coupling $e$, it is still at high energies from the point of view of the NLSM: such a NLSM is not conformal and will continue to flow towards the IR. 

Using the formul\ae{} in Appendix~\ref{Metric and B-field}, one can compute the metric $g_\text{UV}$ and the NS-NS three-form field $H_\text{UV}$ from the generalized potential (\ref{generalized potential conifold}). Since we do not expect $g_\text{UV},H_{UV}$ to solve the supergravity equations,%
\footnote{These can be written as a single differential equation for $K$ \cite{Hull:2010sn}, which in the case of precisely one pair of semichiral fields and a chiral field was given in \cite{Halmagyi:2007ft}.}
we are not particularly interested in their explicit expression, apart from the fact that for finite $\beta$ one finds $H_{UV}\neq 0$ and $J_+\neq J_-$ and therefore the target in the UV is generalized K\"{a}hler. In the special limit $\beta\rightarrow \infty$, the $H$-field vanishes and one recovers the K\"{a}hler case, as expected. The full expression for $H_\text{UV}$ is rather lengthy, but can be computed explicitly.
%To first order in $1/\beta$ one finds 
%\equ{
%H_\text{UV} = -\frac{1}{\beta}\, \frac{4\, \xi|X_L|^2}{(\xi^2+4r^2)^\frac{3}{2}} \, \( \phi \, dX_L\wedge d\bar{X}_R \wedge d\bar \phi+c.c.\) + \cdots  \;,
%}

We now wish to address the issue of instanton corrections. Since  these are independent of the RG scale, one may compute them in the UV NLSM, without knowledge of the IR metric and flux. In the K\"{a}hler case these arise from multi-coverings of the worldsheet onto the blown-up $S^2$, which in our coordinates corresponds to setting $X_R=\phi=0$.

As discussed above, nontrivial solutions to (\ref{generalized holomorphic equations}) can arise only from submanifolds on which $(J_+-J_-)$ is degenerate. We now investigate whether this model contains such submanifolds, of complex dimension one which may harbor nontrivial instanton corrections to $Z_{S^{2}}$. One possibility is the submanifold $X_R=X_L=0$ which, however, in our model is non-compact and therefore cannot harbor nontrivial instanton corrections (moreover, this submanifold has nothing to do with the $S^2$ that hosts the instantons in the $\beta\to\infty$ limit). This seems to be at odds with the  supersymmetric localization result (\ref{instanton corrections conifold}).

Another possibility relates to the interesting phenomenon of type-change in generalized K\"{a}hler manifolds \cite{Gualtieri:2003dx, Cavalcanti:2007, Cavalcanti:2008ur}. This occurs when $J_+ - J_-$ (or $J_+ + J_-$) develops a \textit{new} zero eigenvalue on a certain locus, signaling that on that locus a pair of semichiral fields has turned into a pair of chiral (or twisted chiral) fields (see discussion in Appendix~\ref{Type-change loci}).

Let us investigate the possibility of type-change in the model (\ref{generalized potential conifold}). Using the formula (\ref{eigen chiral appendix}), the eigenvalues of $\frac{1}{2}(J_+ - J_-)$ are $\{0, \pm i\lambda\}$ (each with multiplicity two) with
\equ{
\lambda^2= - \frac{\xi \Big( |X_L|^2 |\phi|^2-|X_R|^2 \Big) - \Big( |X_R|^2 + \big( 2+|X_L|^2 \big) |\phi|^2 \Big) \sqrt{\xi^2+4r^2}}{2r^2\beta^2 \Big( \sqrt{\xi^2+4 r^2}+\frac{1}{\beta} \big( X_L X_R+\bar X_L\bar X_R \big) \Big) } \;.
}
The only locus on which the numerator vanishes is $\phi=X_R=0$. However in this case $r=0$ and the denominator vanishes as well. Taking the limit one finds $\lambda^2\rightarrow \tfrac{1}{\beta^2(1+|X_L|^2)}$. Thus, for finite $\beta$ there are no points in the patch under consideration where $\lambda=0$; a similar analysis in the other patch $\Phi_2\neq 0$ leads to the conclusion that there are no type-change loci for $J_+ - J_-$. One can then check that $ω$ is also non-degenerate since the metric is well defined at the tip.

This result is a bit surprising. On the one hand, the absence of compact submanifolds where $J_+ - J_-$ is degenerate seems to imply that the only finite-action solutions to (\ref{generalized holomorphic equations}) are constant maps, and therefore that the partition function does not receive instanton corrections. This is the expectation, for instance, in \cite{Kapustin:2004gv}. On the other hand, the partition function computed in (\ref{instanton corrections conifold}) seems to represent instanton corrections (which are identical to those of the K\"ahler model in the $\beta \to \infty$ limit). Although we do not have a clear resolution of this puzzle, that we leave as an open question, we propose the following possibility: that the partition function, computed with our contour prescription, captures complexified solutions to the equations (\ref{generalized holomorphic equations}) for the fields in Euclidean signature. Another possibility that we cannot exclude, though, is that---because of the unavoidable divergences---the partition function computed in this paper is not a well-defined object.

\section{Discussion}\label{Discussion}

In this paper we have studied a class of two-dimensional GLSMs with a gauged Wess-Zumino term and off-shell $\N=(2,2)$ SUSY. These involve chiral and semichiral fields coupled to the usual vector multiplet, and are described at low energies by NLSMs on generalized K\"ahler manifolds, as opposed to the more standard case of K\"ahler manifolds when there is no gauged Wess-Zumino term. As we have shown, the GLSMs can be placed on the round $S^2$ (with the untwisted background of \cite{Benini:2012ui, Doroud:2012xw}) while preserving all supercharges \cite{Closset:2014pda}, and we have explicitly constructed their actions. We have also shown that the parameters controlling the gauged Wess-Zumino coupling enter in $\Q_A$-exact action terms. Thus, localization should be insensitive to the non-K\"ahler deformation, which we have verified explicitly. Unfortunately, these theories do not admit enough twisted masses to remove all massless modes, and their partition functions are inherently divergent. We have computed the partition functions by means of supersymmetric localization on the Coulomb branch, and proposed a contour prescription to remove the singularities. 

In principle, the techniques described here provide a method for computing the partition function of NLSMs on certain generalized K\"ahler manifolds, for which currently no other method exists. However, as discussed, this approach raises some puzzles which we have not fully resolved.  Although we discussed possible resolutions, this certainly deserves further study.

As a simple but interesting example illustrating our point we have considered a GLSM realizing a one-parameter family of generalized K\"ahler structures on the conifold. Although the generalized holomorphic equations (\ref{generalized holomorphic equations})  of the A/B-model do not admit real, compact, solutions in this case (apart from  constant maps), the partition function as computed by localization does seem to exhibit instanton contributions. This raises the question of how to reconcile the two statements. We hope that our observation can thrust some progress in the study of these topological models.

We should mention that semichiral fields can be T-dualized to a chiral plus  twisted chiral field \cite{Grisaru:1997ep, Lindstrom:2007sq}; it would be very interesting if supersymmetric localization could shed light into aspects of generalized mirror symmetry, especially non-perturbative ones. Regarding possible extensions, it may be interesting to apply localization techniques to GLSMs realizing generalized K\"{a}hler manifolds without semichiral fields, which are constructed  by coupling  chiral and twisted chiral fields to the Large Vector Multiplet \cite{Lindstrom:2007vc, Lindstrom:2008hx}. These models can realize compact manifolds with an $H$-field as well as noncompact ones.

\section*{\centering Acknowledgements}

We would like to thank Nikolay Bobev, Kentaro Hori, Sungjay Lee, Ulf Lindstr\"{o}m, Guli Lockhart, Daniel Park, Wolfger Peelaers, Martin Roček and Stefan Vandoren for discussions. We especially thank Martin Roček for useful discussions and collaboration on related problems.

F.B. is supported in part by the Royal Society as a Royal Society University Research Fellowship holder. This research was supported in part by the National Science Foundation under Grant No. NSF PHY11-25915. F.B. thanks the KITP and the program ``New Methods in Nonperturbative Quantum Field Theory'' for hospitality.
P.M.C. is supported by the Netherlands Organization for Scientific Research (NWO) under the VICI Grant 680-47-603. This work is part of the D-ITP consortium, a program of the NWO that is funded by the Dutch Ministry of Education, Culture and Science (OCW). P.M.C. would like to thank the ``2014 Summer Simons Workshop in Mathematics and Physics'' at Stony Brook, during which part of this work was done.
D.J. is supported by MOST under the Grant No. 104-2811-M-002-026. A major part of this work was done at YITP, SBU where D.J. and J.N. were supported in part by NSF Grant No. PHY-1316617.

\appendix

\section[\texorpdfstring{$\N=(2,2)$ Supersymmetry}{N=(2,2) Supersymmetry}]{$\boldsymbol{\N=(2,2)}$ Supersymmetry}\label{app: susy}

\subsection{Conventions for Spinors in Euclidean Space}\label{Spinors in Euclidean space}

Here we give our conventions for spinors in Euclidean signature and some useful identities. We use anticommuting Dirac spinors, and contract them as
\equ{
\epsilon \lambda \equiv \epsilon^\alpha \lambda_\alpha \;,\qquad\quad \epsilon \gamma^\mu \lambda \equiv \epsilon^\alpha (\gamma^\mu)_\alpha\,^\beta \lambda_\beta \;,
}
where the spinors have components labelled as $\lambda_\alpha = \big(\begin{smallmatrix} \lambda_+ \\ \lambda_- \end{smallmatrix}\big)$ (we take lower index to denote a ``column vector'') and the gamma matrices with the above index structure read
\equ{
\gamma^1 = \begin{pmatrix} 0 & 1\\ 1 & 0 \end{pmatrix} \;,\qquad \gamma^2 = \begin{pmatrix} 0 & -i\\ i & 0 \end{pmatrix} \;,\qquad \gamma^3 = -i\gamma^1 \gamma^2 = \begin{pmatrix} 1 & 0\\ 0 & -1 \end{pmatrix} \;.
}
The spinor indices can be raised and lowered using the antisymmetric tensors $C^{\alpha\beta}$ and $C_{\alpha\beta}$, respectively, with $C^{+-} = C_{-+} = 1$. For instance:
\equ{
\lambda^\alpha = C^{\alpha\beta} \lambda_\beta \;,\quad \lambda_\alpha = C_{\alpha\beta} \lambda^\beta \qquad\Rightarrow\qquad \lambda^+ = \lambda_- \;,\quad \lambda^- = -\lambda_+ \;.
}

\subsection[\texorpdfstring{Supersymmetry on $\Bbb R^2$}{Supersymmetry on R²}]{Supersymmetry on $\boldsymbol{\pmb{\Bbb R}^2}$}\label{Supersymmetry on $R^{2}$}

The algebra of $\N=(2,2)$ spinor derivatives on flat space is
\equ{
\{ \Bbb D_{\pm}, \bar{\Bbb D}_{\pm} \} = \pm2i \partial_{\pm\pm}
}
where $\partial_{\pm\pm} = \frac{1}{2}\big( \partial_{1}\mp i \partial_{2} \big)$ are spacetime derivatives and $\square = 2\{\partial_{++}, \partial_{--}\}$, while all other anticommutators vanish. They can be written in terms of spinor coordinates $\theta^\pm$, $\bar\theta^\pm$ as 
\equ{
\Bbb D_\pm = \frac{\partial}{\partial \theta^\pm}\pm i \bar{\theta}^\pm \partial_{\pm\pm} \;,\qquad\quad \bar{\Bbb D}_\pm = \frac{\partial}{\partial \bar \theta^\pm}\pm i \theta^\pm \partial_{\pm\pm} \;.
}
We will use the notation
\equ{
M \big| \,\equiv\, M \big|_{\theta^\pm = \bar\theta^\pm = 0}
}
for the bottom component of a multiplet. The SUSY transformations are generated by
\equ{
\label{SUSY generator}
\delta = \bar{\epsilon}\, \Bbb Q+ \epsilon\, \bar{\Bbb Q}= \bar{ \epsilon}^{+} \Bbb Q_{+} + \bar{\epsilon}^- \Bbb Q_{-}+ \epsilon^{+} \bar{\Bbb Q}_{+}+ \epsilon^{-} \bar{\Bbb Q}_{-} \;.
}
We consider $\epsilon$ and $\bar \epsilon$ as two independent anticommuting Dirac spinors. The supercharges satisfy $\{ \Bbb Q_{\pm}, \bar{\Bbb Q}_{\pm} \} = \mp 2i \partial_{\pm}$ and anticommute with the spinor derivatives. In terms of spinor coordinates:
\equ{
\Bbb Q_\pm = \frac{\partial}{\partial \theta^\pm}\mp i \bar{\theta}^\pm \partial_{\pm\pm} \;,\qquad\quad \bar{\Bbb Q}_\pm = \frac{\partial}{\partial \bar \theta^\pm}\mp i \theta^\pm \partial_{\pm\pm} \;.
}

\subsection[\texorpdfstring{$\N=(2,2)$ Supermultiplets}{N=(2,2) Supermultiplets}]{$\boldsymbol{\N=(2,2)}$ Supermultiplets}\label{N=(2,2) Multiplets}

\subsection*{\it Vector Multiplet}

To formulate gauge theories, one introduces a vector multiplet in superspace. There are various such vector multiplets. The most standard one is the vector multiplet $V$, with gauge transformation $\delta_g V= i (\bar \Lambda-\Lambda )$, with $\Lambda$ a chiral gauge parameter. To derive SUSY transformation rules and explicit actions in component form, it is most convenient to introduce gauge-covariant superderivatives $\nabla_\pm$, $\bar\nabla_\pm$. These can be constructed in different representations, according to the matter fields they are acting on. In chiral representation (where all objects transform with a chiral gauge parameter), they are given in terms of the usual superderivatives by
\equ{
\nabla_\pm = e^{-V}\Bbb D_\pm e^V \;,\qquad\quad \bar \nabla_\pm=\bar{\Bbb D}_\pm \;,
}
and they satisfy the algebra
\equ{
 \{ \nabla_\pm, \bar\nabla_\pm \} = \pm2i D_{\pm\pm} \;,\qquad\quad \Sigma= \{ \bar{\nabla}_{+},\nabla_{-}\} \;,
 }
where $D_{\pm\pm}$ denote the gauge-covariant spacetime derivatives while $\Sigma$ is the field strength supermultiplet. In the Abelian case $\Sigma=\bar{\Bbb D}_+ \Bbb D_- V$.

The component fields of the vector multiplet $\Sigma$ are defined by  
\eqs{
\sigma= \Sigma|  \;,\qquad  i \lambda_{+} = \nabla _{+}\Sigma| \;,\qquad -i \lambda_{-} = \nabla _{-}\bar \Sigma| \;,\qquad - i \tilde D = \nabla _{+}\bar \nabla _{-}\Sigma| \;, \\
\bar{ \sigma}= \bar{\Sigma} |  \;,\qquad  i \bar{\lambda}_{+} = \bar{\nabla }_{+}\bar{\Sigma}| \;,\qquad -i \bar{\lambda}_{-}= \bar{\nabla }_{-} \Sigma| \;,\qquad - i \bar{\tilde D} = \bar{\nabla }_{+}\nabla _{-}\bar{\Sigma}| \;, 
}
where we used the complex notation
\equ{
\sigma = i\sigma_1 - \sigma_2 \;,\qquad \bar\sigma = -i\sigma_1 - \sigma_2 \;,\qquad \tilde D = -i F_{12} +D \;,\qquad \bar{\tilde D} = -i F_{12} -D \;,
}
and  $-i F_{12}=[D_{1},D_{2}]=2 i [D_{--},D_{++}]$.

One may similarly define gauge-covariant supercharges $\Q_\pm$, $\bar\Q_\pm$ such that the SUSY transformations are generated by  
\equ{
\delta=\bep\Q+ \e\bar \Q=\bep^+\Q_+ + \bep^-\Q_- + \e^+\bar{\Q}_+ +\e^-\bar{\Q}_- \;.
}

\subsection*{\it Matter Multiplets}

To define the components of $\Bbb X$ and $\bar{\Bbb X}$ we use the gauge-covariant superderivatives: 
\equ{
\label{components X Appx}
\begin{gathered}
X = \Bbb X\big| \;,\qquad\qquad \psi_{\pm}= \,\nabla_{\pm}\Bbb X \big| \;,\qquad\qquad   F= \nabla_{+}\nabla_{-}\Bbb X\big| \;, \\
\bar X=\bar{\Bbb X}\big| \;,\qquad\qquad \bar{\psi}_{\pm}= \,\bar{\nabla}_{\pm}\bar{\Bbb X} \big| \;,\qquad\qquad   \bar F= \bar{\nabla}_{+} \bar{\nabla}_{-}\bar{\Bbb X}\big| \;, \\ 
\bar \chi_{\pm} = \bar{\nabla}_{\pm}\Bbb X\big| \;,\quad  M_{\mp\pm}=\nabla_{\pm}\bar{\nabla}_{\mp} \Bbb X\big| \;,\quad  M_{\pm \pm}=\nabla_{\pm} \bar{\nabla}_{\pm}\Bbb X \big| \;,\quad  \bar \eta_{\pm}= \nabla_{+} \nabla_{-}\bar{\nabla}_{\pm} \Bbb X\big| \;, \\ 
\chi_{\pm} =\nabla_{\pm}\bar{\Bbb X}\big| \;,\quad  \bar M_{\mp\pm}=\bar{\nabla}_{\pm}\nabla_{\mp} \bar{\Bbb X}\big| \;,\quad  \bar M_{\pm \pm}=\bar{\nabla}_{\pm}\nabla_{\pm} \bar{\Bbb X} \big| \;,\quad  \eta_{\pm} = \bar\nabla_+ \bar\nabla_- \nabla_{\pm} \bar{ \Bbb X}\big| \;.
\end{gathered}
}
The multiplet $\Bbb X$ can describe chiral, twisted chiral,%
\footnote{A twisted chiral field cannot couple minimally to the vector multiplet. Thus, in the case of a twisted chiral field, these identifications and the SUSY transformations are valid only when the field is neutral. For instance, the twisted chiral field could be taken to be the Abelian vector multiplet $\Sigma$ with the following identifications: $X=\sigma$, $\psi_+ = i \lambda_+$, $\bar{\chi}_{-} = -i \lambda_{-}$, $M_{-+} = - i \tilde D$.}
as well as left and right semichiral multiplets as follows:
\eqsn{
\text{Chiral}: \quad &\bar \chi_{\pm}=M_{\pm \pm}=M_{\pm\mp}=\bar{\eta}_{\pm}=0 \\
\text{Twisted Chiral}: \quad &\psi_{-}=\bar{\chi}_{+}=F=M_{+\pm}=\bar{\eta}_{+}=0 ,\; M_{--}=-2i D_{--}X ,\; \bar{\eta}_{-}= -2i D_{--}\psi_{+} \\
\text{Left Semichiral}: \quad &\bar \chi_{+}=M_{+\pm}=\bar{\eta}_{+}=0 \\
\text{Right Semichiral}: \quad &\bar \chi_{-}=M_{-\pm}=\bar{\eta}_{-}=0 \;.
}

The most convenient way to determine the SUSY transformations of the component fields, defined by expressions such as $\left[ \nabla^n_\pm \,  \bar{\nabla}^m_\pm \, \Bbb X\right]\big|$ above, is by using identities such as
\equ{
\delta\left[ \nabla^n_\pm \,  \bar{\nabla}^m_\pm \, \Bbb X\right]\big|\equiv \left[ \nabla^n_\pm \,  \bar{\nabla}^m_\pm \, \delta \, \Bbb X\right]\big|  =\left[(\bep \, \nabla+ \e \, \bar \nabla) \nabla^n_\pm \,  \bar{\nabla}^m_\pm \, \Bbb X\right]\big| \;,
}
where we have used the fact that $\nabla_\pm$'s and $\Q_\pm$'s anticommute and that $\nabla_\pm|=\Q_\pm|$, and similarly for the barred operators. From this we find:
\eqsg{
\label{transf rules components flat}
\delta X&= \bep \psi +\e \bar \chi\,,\\
\delta \psi_{+} &= -\bar{\epsilon}^{-}F +\e^{+} 2iD_{++}X+\e^{-}\bar{ \sigma}X  - \e^{+}M_{++} -\e^{-}M_{-+}\,,\\
\delta \psi_{-}&=\bep^{+}F + \e^{+} \sigma X - \e^{-}2iD_{--}X  -\e^{+}M_{+-}-\e^{-}M_{--}\,,\\
\delta F&=\e^{+}2iD_{++} \psi_{-} + \e^{-}2iD_{--}\psi_{+} - \e^{+} \sigma \psi_{+} + \e^{-}\bar{ \sigma} \psi_{-}- i( \e^{+}\lambda_{+}+ \e^{-} \lambda_{-})X +\e \bar \eta\,,\\
\delta \bar \chi_{±} &= \bep_{-}M_{±+} -\bep_{+}M_{±-}\,, \\
\delta M_{±+} &=\bep_{+}\bar \eta_{±}+\e_{-}2i D_{++}\bar \chi_{±}-\e_{+}\bar{ \sigma}\bar \chi_{±}\,, \\
\delta M_{±-} &=\bep_{-}\bar \eta_{±}+\e_{+}2i D_{--}\bar \chi_{±}+\e_{-} \sigma\bar \chi_{±}\,, \\
\delta \bar \eta_{+}&= -i (\e \lambda) \bar \chi_{+}+\e_{+}(-2iD_{--}M_{++}-\bar{ \sigma} M_{+-}) +\e_{-}(2i D_{++}M_{+-}- \sigma M_{++})\,,\\
\delta \bar \eta_{-}&= -i (\e \lambda) \bar \chi_{-}+\e_{-}(2iD_{++}M_{--}- \sigma M_{-+}) +\e_{+}(-2i D_{--}M_{-+}-\bar{ \sigma } M_{--})\,,
}
as well as similar transformations for $\bar{\Bbb X}$. Using identities such as 
\equn{
\bar{\epsilon}_{-}2i D_{++} \psi_{-} -\bar{\epsilon}_{+}2i D_{--} \psi_{+} = i\bar{\epsilon} \gamma^{\mu}D_{\mu}\psi \;,
}
one may write these transformations in the form (\ref{transf rules flat space}). For certain computations (such as proving invariance of the action on $S^{2}$) we find it convenient to keep the index notation and introduce the operator
\equn{
\P_{\alpha \beta}\equiv\pmat{2i D_{++}&  \sigma \\ \bar{ \sigma} & -2i D_{--}} \;,
}
whose supersymmetric variation is $\delta\P_{\alpha\beta}=i\(\epsilon_\alpha\bar{\lambda}_\beta+\bar{\epsilon}_\beta\lambda_\alpha\)$. In this way, all transformation rules can be written in the compact form:
\equ{\begin{split}
\delta X&=\bar{\epsilon} \psi + \epsilon \bar{\chi}\,, \\
\delta\psi_\alpha&=\epsilon^\beta( \P_{\beta \alpha} X -M_{\beta\alpha}) + \bar{\epsilon}_\alpha F\,, \\
\delta F&=\epsilon^\alpha\P_{\alpha \beta} \psi^\beta - i(\epsilon\lambda)X + \epsilon\bar{\eta}\,, \\
\delta \bar{\chi}_{\alpha}&= M_{\alpha \beta}\bep^{\beta}\,, \\
\delta M_{\alpha\beta}&=\epsilon^\gamma\P_{\gamma \beta} \bar{\chi}_\alpha - \bar{\eta}_\alpha\bar{\epsilon}_\beta \,, \\
\delta\bar{\eta}_\alpha &=\epsilon^\kappa\P_{\kappa \gamma} M_{\alpha\beta}C^{\gamma\beta} - i(\epsilon\lambda)\bar{\chi}_\alpha\,,
\end{split}\qquad \qquad
\begin{split}
\delta\bar X&=\epsilon \bar{\psi} + \bar{\epsilon} \chi\,, \\
\delta \bar{\psi}_\alpha &= \bar{\epsilon}^\beta( \P_{\alpha \beta} \bar{X} -\bar{M}_{\beta\alpha}) +\e_\alpha \bar F\,, \\
\delta \bar F&= \bep^\alpha\P_{\beta\alpha}\bar{\psi}^\beta - i (\bar{\epsilon}\bar{\lambda})\bar{X} +\bar \epsilon \eta\,,\\
\delta \chi_{\alpha}&= \bar{ M}_{\alpha \beta}\e^{\beta}\,,\\
\delta \bar{M}_{\alpha \beta}& = \bep^{\gamma}\P_{\beta \gamma}\chi_{\alpha}-\eta_{\alpha}\epsilon_{\beta}\,,\\
\delta \eta_{\alpha}& = \bep^{\kappa}\P_{\gamma \kappa}\bar{M}_{\alpha \beta}C^{\gamma \beta}- i (\bep \bar \lambda)\chi_{\alpha}\,.
\end{split}
\label{flat variation}}
By setting the appropriate fields to zero, these become the SUSY transformations for chiral, twisted chiral and semichiral multiplets.

The component action is found by writing the spinorial measure as
\equ{
\int d^4 \theta \; \bar{\Bbb X} \Bbb X=\left[\nabla_+\nabla_- \bar{ \nabla}_+\bar{ \nabla}_-\, (\bar{\Bbb X} \Bbb X)\right]\Big| \;,
}
using that the rules for Grassmann integration and differentiation are the same, and that the integrand is gauge-invariant.

\subsection*{\it Semichiral Vector Multiplet}

Here we give a brief review of the Semichiral Vector Multiplet (SVM for short), following  \cite{Lindstrom:2007vc, Lindstrom:2008hx}. For simplicity we describe the Abelian case. The SVM gauges isometries that act only on semichiral fields and is defined by four vector multiplets $(V_{L},V_{R}, \Bbb V,\tilde{\Bbb V})$, with gauge transformations
\equ{
\label{gauge transformations SVM}
\delta V_{L}=i(\bar \Lambda_{L} - \Lambda_{L}) \;, \quad  \delta V_{R}=  i (\bar \Lambda_{R} - \Lambda_{R}) \;, \quad  i \delta \Bbb V = i ( \Lambda_{L} - \Lambda_{R}) \;, \quad  i \delta \tilde{ \Bbb V} = i (\Lambda_{L} -\bar \Lambda_{R}) \;,
}
where $\Lambda_{L,R}$ are semichiral fields. These vector multiplets are not independent, but satisfy 
\equ{\label{relations SVM}
-\tfrac{1}{2}V'\equiv \text{Re} \, \Bbb{\tilde V}=\text{Re}\, \Bbb V \,,\qquad \text{Im}\, (\Bbb{\tilde V} -  \Bbb{V})=V_{R}\,,\qquad \text{Im}\,( \Bbb{\tilde V} +  \Bbb{V})=V_{L}\,. 
}
Then, 
\equ{
\label{V and V tilde}
\Bbb V = \tfrac{1}{2} \big( -V'+i (V_L-V_R) \big) \;,\qquad\quad \tilde{ \Bbb V} = \tfrac{1}{2} \big( -V'+i (V_L+V_R) \big) \;,
}
where $V'$ must transform under gauge transformations as $\delta  V' =(\Lambda_R + \bar  \Lambda_R- \Lambda_L -\bar  \Lambda_L)$. There are two field strengths which are invariant under the full gauge symmetry (\ref{gauge transformations SVM}):
\equ{
\label{definition field strengths}
\Bbb F\equiv\bar{\Bbb D}_{+}\bar{\Bbb D}_{-} \Bbb V \qquad\quad\text{and}\qquad\quad \tilde{\Bbb F}\equiv \bar{\Bbb D}_{+}\Bbb D_{-} \tilde{\Bbb V} \;,
} 
chiral and twisted chiral, respectively. The kinetic action for the SVM is given by
\equ{\label{action SVM}
\L_\text{SVM}=- \frac{1}{2e^{2}}\int d^{4}\theta
 \(\bar {\tilde{ \Bbb F}} \tilde{\Bbb F} -\bar {\Bbb F} \Bbb F \) \;,
}
and the FI terms are given by
\equ{
\L_\text{FI} =\(i \, t \int d^{2}\tilde \theta  \, \tilde{\Bbb F}+c.c.\)+ \(i \, s \int d^{2}\theta  \, \Bbb F+c.c.\) \;.
}
From the definitions (\ref{definition field strengths}), the FI terms can also be written as D-terms.

\subsection[\texorpdfstring{Supersymmetry on $S^2$}{Supersymmetry on S²}]{Supersymmetry on $\boldsymbol{S^2}$}\label{Supersymmetry on S^2}

The metric on the round sphere of radius $r$ is
\equn{
ds^{2} = r^{2} (d \theta^{2}+\sin^{2} \theta\, d \varphi^{2}) \;.
}
We consider Killing spinors on $S^{2}$ satisfying
\equ{
\label{killing spinor eqs appendix}
D_{\mu}\e = \tfrac{i}{2r} \gamma_{\mu}\e \;,\qquad\quad D_{\mu}\bep = \tfrac{i}{2r} \gamma_{\mu}\bep \;.
}
The gauge-covariant derivatives on $S^{2}$ read
\eqsg{
\P_{++}=i D_{1}+D_{2}=& \frac{i}{r} \(\partial_{\theta} -\frac{i}{\sin \theta} \partial_{\varphi}\) -\frac{i s}{r}\frac{\cos \theta}{\sin \theta} \;,\\
\P_{--}=-i D_{1}+D_{2}=& -\frac{i}{r} \(\partial_{\theta} +\frac{i}{\sin \theta} \partial_{\varphi}\) -\frac{i s}{r}\frac{\cos \theta}{\sin \theta} \;, 
}
where $s= s_{z} -\frac{\rho(\mathfrak{m})}{2}$ is the effective spin. Using these derivatives, one can check that 
\equ{
\pmat{\e_{+} \\ \e_{-}}= e^{i \frac{\varphi}{2}}\pmat{\cos \frac{\theta}{2} \\ i \sin \frac{\theta}{2} }\,, \qquad \pmat{\bep_{+} \\ \bep_{-}}= e^{-i \frac{\varphi}{2}}\pmat{i \sin \frac{\theta}{2} \\ \cos \frac{\theta}{2} }
\label{expression explicit killing spinor}
}
satisfy the Killing spinor equations (\ref{killing spinor eqs appendix}). We use the latter spinors $\epsilon$, $\bar\epsilon$ to construct the localizing supercharge $\Q_A$.

\subsection{Weyl Covariance}\label{Weyl Covariance}

The additional terms that are supplemented to the $\Bbb{R}^2$ transformations of fields of a given R-charge, are determined by the requirement that the SUSY transformations (\ref{flat variation}) are covariant under Weyl transformations. The metric transforms infinitesimally as $\delta g_{\mu\nu} = 2\Omega\,g_{\mu\nu}$, hence it follows that the spin connection transforms as
\equ{
\delta{\omega_\mu}^{mn}={e_\mu}^m{e_\nu}^n\partial^\nu\Omega \;.
}
For a field $\varphi$ of spin $s_z$ and weight $w$ under a Weyl transformation, $i.e.$, $\delta\varphi = -w \Omega\,\varphi$, we have:
\equ{
\delta(\P_{\pm\pm}\varphi) = -w \Omega\,\P_{\pm\pm}\varphi -(w \pm s_z)  (\P_{\pm\pm}\Omega) \varphi \;.
}
This uniquely determines the additional terms one must add to the flat-space transformation rules (in addition to replacing the derivatives by covariant derivatives) to determine the transformation rules on $S^{2}$. We assume that $\epsilon$ is a ``positive'' Killing spinor satisfying
\equ{
\label{killing spinor appendix}
D_\mu \epsilon=\tfrac{i}{2r}\gamma_\mu \epsilon \qquad\Rightarrow\qquad \P_{\pm\pm}\epsilon^\pm = -\tfrac{1}{r}\epsilon_\pm \;.
}
We also take the spinor $\bep$ to satisfy the same equation. Then the additional terms follow by the replacement rule
\eqs{\label{prescription susy transf S2}
\epsilon^{\pm}\P_{\pm \pm} \varphi \quad \xrightarrow{\text{replace}}\quad & \epsilon^{\pm}\P_{\pm \pm}\varphi \,+\, (-1)^F (w \pm s_z) \varphi\, \P_{\pm \pm} \epsilon^{\pm}\nonumber\\
=&\,\epsilon^{\pm}\P_{\pm \pm}\varphi \,-\, (-1)^F (w \pm s_z)\tfrac{1}{r}\,\varphi\,\epsilon_{\pm} \;,
}
where $F=0$ if $\varphi$ is a boson and $F=1$ if it is a fermion, and we have used the Killing spinor equation (\ref{killing spinor appendix}).

We denote the scaling dimension of the lowest component  $X=\Bbb X|$ by $\frac{q}{2}$. From the definitions (\ref{components X Appx}), Weyl weights and R-charges of all the other component fields are determined and given in Table~\ref{all charges components semi}. Using those charges and following the prescription (\ref{prescription susy transf S2}), we find that the SUSY transformations on $S^{2}$ are given by \eqref{SUSY variations sphere 1}.

\subsection{BPS Equations for the Semichiral Multiplet}\label{BPS equations for the semichiral multiplet}

Here we show that the BPS equations following from the transformation rules (\ref{SUSY variations sphere 1}) have only one smooth solution for $q\neq 0$: $X=F=M=0$. Let us first write down the BPS equations following from $\Q_A \psi_{\pm} = \Q_A \bar{\psi}_{\pm}=0$. Using (\ref{expression explicit killing spinor}) we find
\bea
0 &=-\sin \tfrac{\theta}{2}\( 2D_{++}X- i e^{-i \varphi}F+i M_{++}\)-\cos \tfrac{\theta}{2} \(\bar{{\sigma}}X +\tfrac{q}{2r} X -M_{-+}\),\\
0 &=+\cos \tfrac{\theta}{2}\( 2D_{--}X- i e^{-i \varphi}F-i M_{--}\)+\sin \tfrac{\theta}{2} \({\sigma}X -\tfrac{q}{2r} X -M_{+-}\),\\
0 &=+\cos\tfrac{\theta}{2}\(2D_{++}\bar{X} -i e^{iϕ}\bar{F}+i\bar{M}_{++}\) +\sin\tfrac{\theta}{2}\(\bar{X}{\s} -\tfrac{q}{2r}\bar{X}+\bar{M}_{-+}\),\\
0 &=-\sin\tfrac{\theta}{2}\(2D_{--}\bar{X} -i e^{iϕ}\bar{F} -i\bar{M}_{--}\) -\cos\tfrac{\theta}{2}\(\bar{X}\bar{{\s}} +\tfrac{q}{2r}\bar{X} +\bar{M}_{+-}\).
\eea
Recall that for a particular semichiral field, only some components of $M_{\alpha\beta}$ are non-zero in these equations. Note also that for $M_{\alpha \beta}=0$, these reduce to the BPS equations for a chiral field of R-charge $q$.

Let us now look at the equations following from other spinor variations. Take for instance $\Bbb{X}_R$ and $\bar{\Bbb{X}}_R$, $i.e.$, $\Q_A \bar{\chi}_+ = \Q_A \chi_+ = \Q_A \bar{\eta}_+ = \Q_A \eta_+=0 \,$:
\eqs{0&=M_{++}\cos\tfrac{\theta}{2} - i M_{+-}\sin\tfrac{\theta}{2} \,,\label{BPS M 1}\\
0&=i \bar{M}_{++}\sin\tfrac{\theta}{2} -\bar{M}_{+-}\cos\tfrac{\theta}{2} \,,\label{BPS Mb}\\
0&=\cos\tfrac{\theta}{2} \(2i D_{--} M_{++} +\(\bar{{\s}} +\tfrac{q+2}{2r}\)M_{+-}\) +\sin\tfrac{\theta}{2} \(2 D_{++} M_{+-} +i\({\s} -\tfrac{q}{2r}\)M_{++}\),\label{BPS M 2}\\
0&=\sin\tfrac{\theta}{2} \(2 D_{--}\bar{M}_{++} -i\({\s} +\tfrac{q+2}{2r}\)\bar{M}_{+-}\) +\cos\tfrac{\theta}{2} \(2i D_{++}\bar{M}_{+-} -\(\bar{{\s}} -\tfrac{q}{2r}\)\bar{M}_{++}\).\label{BPS M 3}
}
Consider the equations \eqref{BPS M 1} and \eqref{BPS Mb} and their complex conjugates. Imposing the reality condition $X^\dag = \bar{X}$, $\psi^\dag = \bar{\psi}$, $M^\dag = \bar{M}$, \etc, immediately leads to
\equ{
0=M_{++}\cos \theta=M_{+-}\cos \theta \qquad\Rightarrow\qquad M_{++}=M_{+-}=0 \;,
}
and \eqref{BPS M 2} and \eqref{BPS M 3} are trivially satisfied.%
\footnote{At the special point $\theta=\frac{\pi}{2}$ the fields $M$ need not vanish, but we restrict ourselves to smooth field configurations.}
The same analysis holds for left semichiral fields  and thus one also concludes $M_{--}=M_{-+}=0$. Finally, plugging $M_{\alpha\beta}=0$ into $\Q_A \psi = \Q_A \bar{\psi}=0$ reduces those equations to the BPS equations for a chiral field. As discussed in \cite{Benini:2012ui, Doroud:2012xw}, for generic $q$ the only smooth solution is $X=F=M=0$.

\section{GLSMs for Semichiral Fields}\label{GLSMs for semichiral fields}

\subsection{Kinetic Action and Positivity of the Metric}\label{Kinetic action and positivity of the metric}

Consider a theory with $N_F$ pairs of semichiral fields $(\Bbb X_L^i, \Bbb X_R^i)$ with $i = 1,\dots, N_F$, where the left and right partners transform in the same representation $\fR$ of the gauge and/or flavor group. Then the most general gauge-invariant quadratic kinetic action follows from the superspace Lagrangian
\equ{
\L= \int d^4 \theta \, \(\beta_{ij}\, \bar\bX_L^i \bX_L^j + \gamma_{ij} \, \bar\bX_R^i \bX_R^j - \alpha_{ij}\, \bar\bX_L^i \bX_R^j - \alpha^\dag_{ij}\, \bar\bX_R^i \bX_L^j \) \;,
}
where $\beta$, $\gamma$ are Hermitian matrices while $\alpha$ is a generic complex matrix. By field redefinitions one can set $\beta$ and $\gamma$ to be diagonal with entries $\pm1,0$. Requiring the metric to be positive-definite after having integrated out the auxiliary fields, leads to the following two conditions:
\equ{
(\alpha^\dag)^{-1} \gamma \alpha^{-1} < 0 \qquad\qquad\text{and}\qquad\qquad \beta (\alpha^\dag)^{-1}\gamma \alpha^{-1} \beta - \beta > 0 \;.
}
These force $\beta = \gamma = - \unit$ and by the singular value decomposition theorem we can, by further unitary field redefinitions, reduce $\alpha$ to a diagonal matrix with non-negative entries which now have to satisfy $α_{ii}>1$ as in the case $N_F=1$. Thus, in the presence of multiple semichiral pairs, we can always choose a basis that diagonalizes the quadratic kinetic action.

\subsection{Semichiral-Semichiral Duality}\label{Semichiral-Semichiral Duality}

An interesting feature of semichiral fields is that a pair $(\Bbb X_{L}, \Bbb X_{R})$ in a representation $(\fR, \fR)$ of the gauge and/or flavor group, is ``dual'' to a pair in representation $(\fR, \wb\fR)$ or $(\wb\fR, \fR)$ \cite{Grisaru:1997ep}.   Unlike T-duality, this is simply a change of coordinates which does not change the geometry (one may call this a coordinate duality). To see how this works, consider a pair of semichiral fields in representation  $(\fR, \fR)$ with an action of the form (\ref{flat space lagrangian left and right}).  The idea is to relax the condition of semichirality on $\Bbb X_R$, imposing it by a semichiral Lagrange multiplier $\tilde{\Bbb X}_R$, $i.e.$,
\equ{
\mathcal L = -\int d^{4}\theta \left[\bar{\Bbb{X}}_L  \Bbb{ X}_L +  \bar{\Bbb{ X}}_R  \Bbb{ X}_R +  \alpha \big( \bar{\Bbb{ X}}_L   \Bbb{ X}_R+ \bar{\Bbb{ X}}_R  \Bbb{ X}_L \big)  - \big( \tilde{ \Bbb X}_R \Bbb X_R + \bar{\tilde{ \Bbb X}}_R \bar{ \Bbb X}_R \big) \right] \;.
}
This action is gauge-invariant provided $\tilde{ \Bbb X}_{R}$ is in representation $\wb\fR$. Integrating out $\tilde{\Bbb X}_{R}$ simply imposes that $\Bbb X_{R}$ is right semichiral and  leads to the original model. On the other hand, integrating out $\Bbb X_R$ leads to the change of coordinates $\tilde{\Bbb X}_R=\bar{\Bbb X}_R+\alpha \, \bar{\Bbb X}_L $ and the dual action reads
\equ{
\mathcal L = \int d^{4}\theta \Big[  (\alpha^{2}-1) \bar{\Bbb{ X}}_L  \Bbb{ X}_L+\bar{\tilde {\Bbb{X}}}_R   \tilde {\Bbb{ X}}_R   - \alpha \big( \tilde{\Bbb{ X}}_R \Bbb{ X}_L+ \bar{\Bbb{ X}}_L \bar{\tilde { \Bbb{ X}}}_R \big) \Big] \;.
\label{action opp charges duality}
}
This is a GLSM for semichiral fields in representation $(\fR, \wb\fR)$. After a rescaling of $\Bbb X_L$ to normalize the first term, we see that the relation between $\alpha$ in (\ref{flat space lagrangian left and right}) and $\beta$ in (\ref{flat space lagrangian left and right opposite charges}) is
\equ{
\beta = \frac\alpha{\sqrt{\alpha^2 - 1}} \;.
}
Of course, one could similarly relax the semichirality condition on $\Bbb X_{L}$ instead, which would lead to a model with semichiral fields in representation $(\wb\fR, \fR)$.

\section{One-loop Determinants}\label{1-loop Determinants}

We first evaluate the bosonic determinant. In addition to the lowest components $X_{L}$ and $X_{R}$ we also have the four fields $M_{\alpha \beta}$ and two auxiliary fields $F_L$, $F_R$, therefore the quadratic terms read $\bar{\X}\O_{B}\X$ with $\bar{\X}=(\bar{X}^L,\bar{X}^R,\bar{M}^R_{+-},\bar{M}^R_{++},\bar{M}^L_{--},\bar{M}^L_{-+}, \bar{F}^L, \bar{F}^R)$ and $\X=(X^L,X^R,M^L_{-+},$ $M^L_{--},M^R_{++},M^R_{+-}, F^L, F^R)^\sT$. The $8\times 8$ matrix of the kinetic operator $\O_{B}$ is then given by
\equ{\pmat{
\O_X & \alpha\,\O_X & \tfrac{q}{2r} -{\s} & 2 i D_{++} & -\alpha\, 2 i D_{--} & -\alpha\(\tfrac{q}{2r} + \bar{{\s}}\) & 0 & 0\\
\alpha\,\O_X & \O_X & \alpha\(\tfrac{q}{2r} -{\s}\) & \alpha\, 2 i D_{++} & -2 i D_{--} &  -\(\tfrac{q}{2r} +\bar{{\s}}\) & 0 & 0\\
\alpha\(-\tfrac{q}{2r} +{\s}\) & -\tfrac{q}{2r} + {\s} & 0 & 0 & 0& -1 & 0 & 0\\
\alpha\, 2 i D_{--} & 2 i D_{--} & 0 & \alpha & 0 & 0 & 0 & 0\\
-2 i D_{++} &  - \alpha \, 2 i D_{++} & 0 & 0 & \alpha & 0 & 0 & 0\\
\tfrac{q}{2r} +\bar{{\s}} & \alpha \(\tfrac{q}{2r} +\bar{{\s}}\)& -1 & 0 & 0& 0 & 0 & 0\\
0 & 0 & 0 & 0 & 0 & 0 & 1 & \alpha \\
0 & 0 & 0 & 0 & 0 & 0 & \alpha & 1}
\label{OXMF appendix}}
where $\O_X$ is given in (\ref{operator O_X}). Expanding the fields in $\X$ in terms of the spherical harmonics $\big(Y_{j,\, j_3}^s,Y_{j,\, j_3}^s,Y_{j,\, j_3}^s,Y_{j,\, j_3}^{s-1},Y_{j,\, j_3}^{s+1},Y_{j,\, j_3}^s, Y_{j,\, j_3}^s, Y_{j,\, j_3}^s \big)$ and using
\equ{
D_{\pm\pm} Y_{j,\, j_3}^s =\pm\tfrac{s_\pm}{2r} Y_{j,\, j_3}^{s\pm 1} \qquad\quad \text{with }\, s_\pm = \sqrt{j (j+1) - s (s \pm 1)} \;,
}
we obtain for $j \geq \frac{|\rho(\fm)|}{2} + 1$:
\begin{multline}
\det \O_{B} = \tfrac{\(\alpha^2 - 1\)^2}{r^4} \left[ j^2 + \tfrac{(\alpha^2 - 1)\rho(\fm)^2}{4} -\alpha^2 \( \tfrac{q}{2} -i r \rho(\sigma_1) \)^2 \right] \times \\
\times \left[(j+1)^2+ \tfrac{(\alpha^2 - 1)\rho(\fm)^2}{4} -\alpha^2 \( \tfrac{q}{2} -i r \rho(\sigma_1) \)^2 \right],
\end{multline}
with multiplicity $2j+1$. Here $\rho(·)$ denotes a weight vector in the representation $\fR$. There are three more cases that need to be considered before we can write down the full determinant.
\begin{itemize}
\item For $j = \frac{|\rho(\fm)|}{2} \geq \frac{1}{2}$, either $Y_{j,\, j_3}^{s+1}$ or $Y_{j,\, j_3}^{s-1}$ does not exist. For instance, for $\rho(\fm) \geq 1$, $Y_{j,\, j_3}^{s-1}$ does not exist, then we can remove the fourth row/column of the matrix $\O_{B}$. Similiarly, for $\rho(\fm) \leq -1$, $Y_{j,\, j_3}^{s+1}$ does not exist, then we can remove the fifth row/column of the matrix $\O_{B}$. In this case, the determinant becomes $\frac{1}{\alpha} \det\O_{B}$ at $j = \frac{|\rho(\fm)|}{2}$.
\item For $j = \rho(\fm) = 0$, only $Y_{0,\, j_3}^0$ exists, and we remove both the fourth and the fifth row/column of the matrix $\O_{B}$. The determinant becomes $\frac{1}{\alpha^2} \det \O_{B}$ at $j = \rho(\fm) = 0$.
\item For $j = \frac{|\rho(\fm)|}{2} - 1 \geq 0$, only one of $Y_{j,\, j_3}^{s+1}$ and $Y_{j,\, j_3}^{s-1}$ exists. In both cases, the determinant is only $\alpha$ with multiplicity $|\rho(\fm)| - 1$.
\end{itemize}
Putting all these cases together, we obtain the full determinant in the bosonic sector (ignoring overall constants):
\begin{multline}
\Det \O_B =\prod_{\rho\in \fR} \frac{ \alpha^{|\rho(\fm)|-1} }{ \alpha^{|\rho(\fm)|+1} } \prod_{j=\frac{|\rho(\fm)|}{2}}^\infty \bigg[ j^2 + \frac{(\alpha^2 - 1) \rho(\fm)^2}4 -\alpha^2 \Big( \frac{q}{2} -i r \rho(\sigma_1) \Big)^2 \bigg]^{2j+1} \times \\
\times \bigg[(j+1)^2 + \frac{(\alpha^2 - 1) \rho(\fm)^2}4 -\alpha^2 \Big( \frac{q}{2} -i r \rho(\sigma_1)\Big)^2 \bigg]^{2j+1} \bigg( \frac{\(\alpha^2 -1\)^2}{r^4} \bigg)^{2j+1} \;.
\end{multline}

Now let us work out the fermionic determinant. First we work out the determinant factor produced by the fields $\psi$ and $\eta$. Proceeding as before, we write the action as $\bar{\Psi}\O_{F}\Psi$ with $\bar{\Psi}=\(\bar{\psi}^{L-},\bar{\psi}^{L+},\bar{\psi}^{R-},\bar{\psi}^{R+},\eta^{R-},\eta^{L+},\chi^{R-},\chi^{L+}\)$, $\Psi=(\psi^{L+},\psi^{L-},\psi^{R+},\psi^{R-},\bar{\eta}^{L+},\bar{\eta}^{R-},\bar{\chi}^{L+},\bar{\chi}^{R-})^\sT$ and $\O_{F}$ is given by
\equ{
\pmat{
-\(\tfrac{q}{2r}+\bar{{\s}}\) & 2 i D_{--} & -\alpha\(\tfrac{q}{2r}+\bar{{\s}}\) & 2 i \alpha D_{--} & -1 & 0 & 0 & 0\\
-2 i D_{++} & \tfrac{q}{2r}-{\s} & -2 i \alpha D_{++} & \alpha\(\tfrac{q}{2r}-{\s}\) & 0 & \alpha & 0 & 0\\
-\alpha\(\tfrac{q}{2r}+\bar{{\s}}\) & 2 i \alpha D_{--} & -\(\tfrac{q}{2r}+\bar{{\s}}\) & 2 i D_{--} & -\alpha & 0 & 0 & 0\\
-2 i \alpha D_{++} & \alpha\(\tfrac{q}{2r}-{\s}\) & -2 i D_{++} & \tfrac{q}{2r}-{\s} & 0 & 1 & 0 & 0\\
-\alpha & 0 & -1 & 0 & 0 & 0 & 0 & 0\\
0 & 1 & 0 & \alpha & 0 & 0 & 0 & 0\\
0 & 0 & 0 & 0 & 0 & 0 & -\alpha\(\tfrac{q}{2r} - {\s}\) & -2 i D_{--}\\
0 & 0 & 0 & 0 & 0 & 0 & 2 i D_{++} & \alpha\(\frac{q}{2r} + \bar{{\s}}\)}.
}
Expanding $\Psi$ in the spherical harmonics $\big(Y_{j,\, j_3}^{s-\frac{1}{2}},Y_{j,\, j_3}^{s+\frac{1}{2}},Y_{j,\, j_3}^{s-\frac{1}{2}},Y_{j,\, j_3}^{s+\frac{1}{2}},Y_{j,\, j_3}^{s-\frac{1}{2}},Y_{j,\, j_3}^{s+\frac{1}{2}},Y_{j,\, j_3}^{s-\frac{1}{2}}$, $Y_{j,\, j_3}^{s+\frac{1}{2}}\big)$, we obtain for $j \geq \frac{|\rho(\fm)|}{2} + \frac{1}{2}$:
\equ{
\det \O_F = \frac{\(\alpha^2 - 1\)^2}{r^4} \left[ \left(j+\frac{1}{2}\right)^2 + \frac{ (\alpha^2 - 1) \rho(\fm)^2}4 -\alpha^2 \Big( \frac{q}{2} -i r \rho(\sigma_1)\Big)^2 \right]^2 \;,
}
with multiplicity $2j+1$. For $j=\frac{|\rho(\fm)|}{2} -\frac{1}{2}\geq 0$ only one of the eigenfunctions exists. For instance, when $\rho(\fm) \geq 1$, only $Y_{j,\, j_3}^{s+\frac{1}{2}}$ exists, which leads to the following eigenvalue with multiplicity $|\rho(\fm)|$:
$$
-\frac{\(\alpha^2-1\)\alpha^2}{r^2} \left( \frac{|\rho(\fm)|}{2}+\frac{q}{2}-i r \rho(\s_1) \right) \left( \frac{|\rho(\fm)|}{2}-\frac{q}{2}+i r \rho(\s_1)\right) \;.
$$
For $\rho(\fm) \leq -1$ only $Y_{j,\, j_3}^{s-\frac{1}{2}}$ exists, and the eigenvalue is same as above.
The case $\rho(\fm) = 0$ does not exist but formally we can still use the same expression. Putting all these cases together, we obtain the full determinant in the fermionic sector:
\begin{multline}
\Det \O_F =\prod_{\rho\in \fR} \bigg( \frac{(1-\alpha^2) \alpha^2}{r^2} \bigg)^{|\rho(\fm)|} \bigg[ \frac{\rho(\fm)^2}4 - \Big( \frac{q}{2}-i r \rho(\s_1)\Big)^2 \bigg]^{|\rho(\fm)|} \times \\
\times \prod_{j=\frac{|\rho(\fm)|+1}2}^\infty \bigg( \frac{\alpha^2 - 1}{r^2}\bigg)^{4j+2} \left[ \left( j+\frac{1}{2}\right)^2 + \frac{(\alpha^2 - 1) \rho(\fm)^2}4 -\alpha^2 \Big( \frac{q}{2} -i r \rho(\sigma_1) \Big)^2 \right]^{4j+2} \;.
\end{multline}
Putting these bosonic and fermionic determinants together leads to many  cancellations and ignoring overall factors, the one-loop determinant  reads
\equ{
\label{one loop det semis appendix}
Z_{LR} =\frac{\text{Det}\,\O_{F}}{\text{Det}\,\O_{B}}= \prod_{\rho \in \fR} \frac{(-1)^{|\rho(\fm)|}}{\frac{\rho(\fm)^2}{4}-\(\frac{q}{2}-i r \rho(\s_1)\)^2} \;.
}

Now we show that  (\ref{one loop det semis appendix}) can be rewritten in a more recognizable form, as a one-loop determinant for chiral fields. For a pair of chiral fields with opposite R-charges and gauge charges one has the expression in (\ref{semichiralPF same charges}) \cite{Benini:2012ui, Doroud:2012xw}. Let us set $q = 0$ for now and then shift $r \rho(\sigma_1)$ to $r \rho(\sigma_1) + i \frac{q}{2}$ in the final result. Using the property of the $\Gamma$-function, $\Gamma(z + n) = (z)_n \cdot \Gamma(z)$, where $(z)_n \equiv \prod_{k = 0}^{n - 1} (z + k)$ is the Pochhammer symbol, we obtain for $\rho(\fm) \geq 1$ (the cases $\rho(\fm) = 0$ and $\rho(\fm) \leq -1$ are similar)
\equn{\prod_{\rho \in \fR} \tfrac{\Gamma \left(-i r \rho(\sigma_1) - \frac{\rho(\fm)}{2} \right)}{\Gamma \left(1 + i r \rho(\sigma_1) - \frac{\rho(\fm)}{2} \right)} \, \tfrac{\Gamma \left(i r \rho(\sigma_1) + \frac{\rho(\fm)}{2} \right)}{\Gamma \left(1 - i r \rho(\sigma_1) + \frac{\rho(\fm)}{2} \right)}
= \prod_{\rho \in \fR} \tfrac{\left(1 + i r \rho(\sigma_1) - \frac{\rho(\fm)}{2}\right)_{\rho(\fm)-1}}{\left(-i r \rho(\sigma_1) - \frac{\rho(\fm)}{2}\right)_{\rho(\fm) + 1}}
= \prod_{\rho \in \fR} \frac{(-1)^{\rho(\fm)}}{\frac{\rho(\fm)^2}{4}-(-i r \rho(\sigma_1))^2} \;,
}
which coincides with (\ref{one loop det semis appendix}) after the appropriate shift.

\section{Target Space Geometry}

\subsection[\texorpdfstring{Metric, $B$-field and Complex Structures}{Metric, B-field and Complex Structures}]{Metric, $\boldsymbol{B}$-field and Complex Structures}\label{Metric and B-field}

For the reader's convenience, here we give some relevant formul\ae{} to compute the metric, the $B$-field and the complex structures $J_\pm$ from the generalized K\"{a}hler potential $K$. For a comprehensive review and details see \cite{Lindstrom:2005zr}. Defining $E=\frac12(g+B)$ one has
\bea
E_{LL} &= C_{LL}K_{LR}^{-1}J_sK_{RL} &
E_{cL} &= C_{cL}K_{LR}^{-1}J_s K_{RL} \\
E_{LR} &= J_sK_{LR}J_s + C_{LL}K_{LR}^{-1}C_{RR} &
E_{cR} &= J_c K_{cR} J_s + C_{cL}K_{LR}^{-1}C_{RR} \\
E_{Lc} &= K_{Lc} + J_s K_{Lc} J_c + C_{LL}K_{LR}^{-1}C_{Rc} &
E_{cc} &= K_{cc}+J_c K_{cc} J_c + C_{cL}K_{LR}^{-1}C_{Rc} \\
E_{Lt} &= -K_{Lt} - J_s K_{Lt} J_t + C_{LL}K_{LR}^{-1}A_{Rt} \qquad &
E_{ct} &= -K_{ct}-J_c K_{ct}J_t + C_{cL}K_{LR}^{-1}A_{Rt} \\
E_{RL} &= -K_{RL}J_s K_{LR}^{-1} J_s K_{RL} &
E_{tL} &= C_{tL}K_{LR}^{-1}J_s K_{RL} \\
E_{RR} &= -K_{RL}J_s K_{LR}^{-1} C_{RR} &
E_{tR} &= J_t K_{tR} J_s + C_{tL}K_{LR}^{-1}C_{RR} \\
E_{Rc} &= K_{Rc} - K_{RL}J_s K_{LR}^{-1} C_{Rc} &
E_{tc} &= K_{tc} + J_t K_{tc} J_c + C_{tL}K_{LR}^{1}C_{Rc} \\
E_{Rt} &= -K_{Rt} - K_{RL}J_s K_{LR}^{-1} A_{Rt} &
E_{tt} &= -K_{tt} - J_t K_{tt} J_t + C_{tL} K_{LR}^{-1} A_{Rt}\,.
\eea
Here $A$ and $C$ are matrices defined as follows (with the two indices suppressed)
\equ{A=\pmat{2iK & 0 \\ 0 & -2iK},\qquad C=\pmat{0 & 2iK \\ -2iK & 0},
}
where $K$ itself is a matrix whose entries are second derivatives of the generalized potential with $c, t, s$ denoting chiral, twisted chiral and semichiral directions, respectively. For instance
\equ{\label{definition KLR}
K_{LR}\equiv\pmat{\frac{\partial^2 K}{\partial X_L \partial X_R} & \frac{\partial^2 K}{\partial X_L \partial \bar{X}_R} \\ \frac{\partial^2 K}{\partial \bar{X}_L \partial X_R} & \frac{\partial^2 K}{\partial \bar{X}_L \partial \bar{X}_R}} \;,
}
and similarly for $K_{Rc}=\frac{\partial^{2}K}{\partial X_{R}\partial \phi}$, \etc 

The complex structures read \cite{Bogaerts:1999jc, Lindstrom:2005zr}
\equ{\label{j+ app}
J_{+}=
\left(\begin{array}{cccc}
J_s &0&0&0\cr
K_{RL}^{-1}C_{LL} &  K_{RL}^{-1}J_s K_{LR} &
K_{RL}^{-1}C_{Lc}
& K_{RL}^{-1}C_{Lt}\cr
0&0&J_c&0\cr
0&0&0&J_t\end{array}\right)~
}
and
\equ{ \label{j- app}
J_{-}=\left(\begin{array}{cccc}
K_{LR}^{-1}J_s K_{RL} & K_{LR}^{-1}C_{RR} &
K_{LR}^{-1}C_{Rc}& K_{LR}^{-1}A_{Rt}\cr
0& J_s&0&0\cr
0&0&J_c&0\cr
0&0&0& -J_t\end{array}\right)\,,
}
where $J_{c,t,s}$ are the canonical complex structures of the form $\spmat{i & 0 \\ 0 & -i}$ and of the appropriate dimension.

\subsection{Type-change Loci}\label{Type-change loci}

The \textit{type} of a generalized  K\"{a}hler structure is given by
\equn{
(k_+,k_-) = \big( \text{dim}_{\Bbb C} \,  \text{ker} (J_+- J_-) \;,\; \text{dim}_{\Bbb C} \,  \text{ker} (J_++ J_-) \big) \;.
}
In terms of $\N=(2,2)$ multiplets, $k_+$ and $k_-$ simply count the number of chiral and twisted chiral fields, respectively; the number of semichiral fields is given by
$$
\text{dim}_{\Bbb C} \,\text{coIm} [J_+, J_-]=d-k_+-k_- \;,
$$
where $\text{coIm}$ is the co-image, while $d$ the complex dimension of the manifold. An important aspect of generalized complex geometry is that the type may jump discontinuously on so-called type-changing loci \cite{Gualtieri:2007ng}.  On these loci, a pair of semichiral fields $(\Bbb X_L, \Bbb X_R)$ becomes either a pair of chiral or a pair of twisted chiral fields.

To study this phenomenon one may compute the eigenvalues of $J_+\pm J_-$. For concreteness, let us consider an arbitrary generalized potential $K$ that depends on a single chiral field and a pair of semichiral fields. At generic points on the manifold the type is $(k_+,k_-)=(1,0)$. From (\ref{j+ app}) and (\ref{j- app}) one finds that the eigenvalues of $\tfrac12(J_+-J_-)$ are $\{0,\pm i \lambda\}$, each with multiplicity two, with 
\equ{
\label{eigen chiral appendix}
\lambda = \sqrt{ \frac{|K_{\bar l r}|^2-K_{r\bar r} K_{l\bar l}}{|K_{\bar l r}|^2-|K_{l r}|^2} } \;.
}
The eigenvalues of $\tfrac12(J_+ + J_-)$ are $\{\pm i,\pm i \tilde \lambda\}$, each with multiplicity two, with 
\equ{
\label{eigen t chiral appendix}
\tilde \lambda = \sqrt{ \frac{|K_{l r}|^2-K_{r\bar r} K_{l\bar l}}{|K_{l r}|^2-|K_{\bar l r}|^2} } \;.
}
On the locus $\lambda=0$ the type jumps to $(k_+,k_-)=(3,0)$, where the manifold is locally described by three chiral fields, and on the locus $\tilde \lambda=0$ the type jumps to $(k_+,k_-)=(1,2)$, where it is locally described by one chiral field and two twisted chiral fields. Whether these loci exist or not depends on the specific potential $K$. For a study of type-change in various WZW models see, for example, \cite{Sevrin:2011mc}.

\bibliographystyle{utphys}
\bibliography{References}

\providecommand{\href}[2]{#2}\begingroup\raggedright\begin{thebibliography}{10}

\bibitem{Pestun:2007rz}
V.~Pestun, ``{Localization of Gauge Theory on a Four-sphere and Supersymmetric
  Wilson Loops},'' \href{http://dx.doi.org/10.1007/s00220-012-1485-0}{{\em
  Commun.Math.Phys.} {\bf 313} (2012)  71--129},
\href{http://arxiv.org/abs/0712.2824}{{\tt arXiv:0712.2824 [hep-th]}}.
%%CITATION = ARXIV:0712.2824;%%.

\bibitem{Witten:1988xj}
E.~Witten, ``{Topological Sigma Models},''
\href{http://dx.doi.org/10.1007/BF01466725}{{\em Commun.Math.Phys.} {\bf 118}
  (1988)  411--449}.
%%CITATION = CMPHA,118,411;%%.

\bibitem{Witten:1991zz}
E.~Witten, ``{Mirror Manifolds and Topological Field Theory},''
  \href{http://arxiv.org/abs/hep-th/9112056}{{\tt arXiv:hep-th/9112056}}.

\bibitem{Benini:2012ui}
F.~Benini and S.~Cremonesi, ``{Partition Functions of ${\mathcal{N}{=}(2,2)}$
  Gauge Theories on $S^2$ and Vortices},''
  \href{http://dx.doi.org/10.1007/s00220-014-2112-z}{{\em Commun.Math.Phys.}
  {\bf 334} (2015)  1483--1527},
\href{http://arxiv.org/abs/1206.2356}{{\tt arXiv:1206.2356 [hep-th]}}.
%%CITATION = ARXIV:1206.2356;%%.

\bibitem{Doroud:2012xw}
N.~Doroud, J.~Gomis, B.~Le~Floch, and S.~Lee, ``{Exact Results in $D=2$
  Supersymmetric Gauge Theories},''
  \href{http://dx.doi.org/10.1007/JHEP05(2013)093}{{\em JHEP} {\bf 1305} (2013)
   093},
\href{http://arxiv.org/abs/1206.2606}{{\tt arXiv:1206.2606 [hep-th]}}.
%%CITATION = ARXIV:1206.2606;%%.

\bibitem{Gomis:2012wy}
J.~Gomis and S.~Lee, ``{Exact Kähler Potential from Gauge Theory and Mirror
  Symmetry},'' \href{http://dx.doi.org/10.1007/JHEP04(2013)019}{{\em JHEP} {\bf
  1304} (2013)  019},
\href{http://arxiv.org/abs/1210.6022}{{\tt arXiv:1210.6022 [hep-th]}}.
%%CITATION = ARXIV:1210.6022;%%.

\bibitem{Gadde:2013dda}
A.~Gadde and S.~Gukov, ``{2d Index and Surface Operators},''
  \href{http://dx.doi.org/10.1007/JHEP03(2014)080}{{\em JHEP} {\bf 1403} (2014)
   080},
\href{http://arxiv.org/abs/1305.0266}{{\tt arXiv:1305.0266 [hep-th]}}.
%%CITATION = ARXIV:1305.0266;%%.

\bibitem{Benini:2013nda}
F.~Benini, R.~Eager, K.~Hori, and Y.~Tachikawa, ``{Elliptic Genera of
  Two-dimensional $\mathcal{N}{=}2$ Gauge Theories with Rank-one Gauge
  Groups},'' \href{http://dx.doi.org/10.1007/s11005-013-0673-y}{{\em
  Lett.Math.Phys.} {\bf 104} (2014)  465--493},
\href{http://arxiv.org/abs/1305.0533}{{\tt arXiv:1305.0533 [hep-th]}}.
%%CITATION = ARXIV:1305.0533;%%.

\bibitem{Sugishita:2013jca}
S.~Sugishita and S.~Terashima, ``{Exact Results in Supersymmetric Field
  Theories on Manifolds with Boundaries},''
  \href{http://dx.doi.org/10.1007/JHEP11(2013)021}{{\em JHEP} {\bf 1311} (2013)
   021},
\href{http://arxiv.org/abs/1308.1973}{{\tt arXiv:1308.1973 [hep-th]}}.
%%CITATION = ARXIV:1308.1973;%%.

\bibitem{Honda:2013uca}
D.~Honda and T.~Okuda, ``{Exact Results for Boundaries and Domain Walls in 2d
  Supersymmetric Theories},''
  \href{http://dx.doi.org/10.1007/JHEP09(2015)140}{{\em JHEP} {\bf 1509} (2015)
   140},
\href{http://arxiv.org/abs/1308.2217}{{\tt arXiv:1308.2217 [hep-th]}}.
%%CITATION = ARXIV:1308.2217;%%.

\bibitem{Hori:2013ika}
K.~Hori and M.~Romo, ``{Exact Results In Two-Dimensional $(2,2)$ Supersymmetric
  Gauge Theories With Boundary},''
\href{http://arxiv.org/abs/1308.2438}{{\tt arXiv:1308.2438 [hep-th]}}.
%%CITATION = ARXIV:1308.2438;%%.

\bibitem{Benini:2013xpa}
F.~Benini, R.~Eager, K.~Hori, and Y.~Tachikawa, ``{Elliptic Genera of 2d
  $\mathcal{N}{=} 2$ Gauge Theories},''
  \href{http://dx.doi.org/10.1007/s00220-014-2210-y}{{\em Commun.Math.Phys.}
  {\bf 333} (2015)  1241--1286},
\href{http://arxiv.org/abs/1308.4896}{{\tt arXiv:1308.4896 [hep-th]}}.
%%CITATION = ARXIV:1308.4896;%%.

\bibitem{Doroud:2013pka}
N.~Doroud and J.~Gomis, ``{Gauge Theory Dynamics and K\"ahler Potential for
  Calabi-Yau Complex Moduli},''
  \href{http://dx.doi.org/10.1007/JHEP12(2013)099}{{\em JHEP} {\bf 1312} (2013)
   099},
\href{http://arxiv.org/abs/1309.2305}{{\tt arXiv:1309.2305 [hep-th]}}.
%%CITATION = ARXIV:1309.2305;%%.

\bibitem{Kim:2013ola}
H.~Kim, S.~Lee, and P.~Yi, ``{Exact Partition Functions on $\mathbb{RP}^2$ and
  Orientifolds},'' \href{http://dx.doi.org/10.1007/JHEP02(2014)103}{{\em JHEP}
  {\bf 1402} (2014)  103},
\href{http://arxiv.org/abs/1310.4505}{{\tt arXiv:1310.4505 [hep-th]}}.
%%CITATION = ARXIV:1310.4505;%%.

\bibitem{Murthy:2013mya}
S.~Murthy, ``{A Holomorphic Anomaly in the Elliptic Genus},''
  \href{http://dx.doi.org/10.1007/JHEP06(2014)165}{{\em JHEP} {\bf 1406} (2014)
   165},
\href{http://arxiv.org/abs/1311.0918}{{\tt arXiv:1311.0918 [hep-th]}}.
%%CITATION = ARXIV:1311.0918;%%.

\bibitem{Ashok:2013pya}
S.~K. Ashok, N.~Doroud, and J.~Troost, ``{Localization and Real Jacobi
  Forms},'' \href{http://dx.doi.org/10.1007/JHEP04(2014)119}{{\em JHEP} {\bf
  1404} (2014)  119},
\href{http://arxiv.org/abs/1311.1110}{{\tt arXiv:1311.1110 [hep-th]}}.
%%CITATION = ARXIV:1311.1110;%%.

\bibitem{Benini:2014mia}
F.~Benini, D.~S. Park, and P.~Zhao, ``{Cluster Algebras from Dualities of 2d
  ${\mathcal{N}}=(2,2)$ Quiver Gauge Theories},''
  \href{http://dx.doi.org/10.1007/s00220-015-2452-3}{{\em Commun.Math.Phys.}
  {\bf 340} (2015) no.~1, 47--104},
\href{http://arxiv.org/abs/1406.2699}{{\tt arXiv:1406.2699 [hep-th]}}.
%%CITATION = ARXIV:1406.2699;%%.

\bibitem{Gomis:2014eya}
J.~Gomis and B.~Le~Floch, ``{M2-brane Surface Operators and Gauge Theory
  Dualities in Toda},''
\href{http://arxiv.org/abs/1407.1852}{{\tt arXiv:1407.1852 [hep-th]}}.
%%CITATION = ARXIV:1407.1852;%%.

\bibitem{Benini:2015noa}
F.~Benini and A.~Zaffaroni, ``{A topologically Twisted Index for
  Three-dimensional Supersymmetric Theories},''
  \href{http://dx.doi.org/10.1007/JHEP07(2015)127}{{\em JHEP} {\bf 1507} (2015)
   127},
\href{http://arxiv.org/abs/1504.03698}{{\tt arXiv:1504.03698 [hep-th]}}.
%%CITATION = ARXIV:1504.03698;%%.

\bibitem{Closset:2015rna}
C.~Closset, S.~Cremonesi, and D.~S. Park, ``{The Equivariant A-twist and Gauged
  Linear Sigma Models on the Two-sphere},''
  \href{http://dx.doi.org/10.1007/JHEP06(2015)076}{{\em JHEP} {\bf 1506} (2015)
   076},
\href{http://arxiv.org/abs/1504.06308}{{\tt arXiv:1504.06308 [hep-th]}}.
%%CITATION = ARXIV:1504.06308;%%.

\bibitem{Witten:1993yc}
E.~Witten, ``{Phases of N=2 Theories in Two-dimensions},''
  \href{http://dx.doi.org/10.1016/0550-3213(93)90033-L}{{\em Nucl.Phys.} {\bf
  B403} (1993)  159--222}, \href{http://arxiv.org/abs/hep-th/9301042}{{\tt
  arXiv:hep-th/9301042}}.

\bibitem{Jockers:2012dk}
H.~Jockers, V.~Kumar, J.~M. Lapan, D.~R. Morrison, and M.~Romo, ``{Two-Sphere
  Partition Functions and Gromov-Witten Invariants},''
  \href{http://dx.doi.org/10.1007/s00220-013-1874-z}{{\em Commun.Math.Phys.}
  {\bf 325} (2014)  1139--1170},
\href{http://arxiv.org/abs/1208.6244}{{\tt arXiv:1208.6244 [hep-th]}}.
%%CITATION = ARXIV:1208.6244;%%.

\bibitem{Honma:2013hma}
Y.~Honma and M.~Manabe, ``{Exact Kähler Potential for Calabi-Yau Fourfolds},''
  \href{http://dx.doi.org/10.1007/JHEP05(2013)102}{{\em JHEP} {\bf 1305} (2013)
   102},
\href{http://arxiv.org/abs/1302.3760}{{\tt arXiv:1302.3760 [hep-th]}}.
%%CITATION = ARXIV:1302.3760;%%.

\bibitem{Park:2012nn}
D.~S. Park and J.~Song, ``{The Seiberg-Witten Kähler Potential as a Two-Sphere
  Partition Function},'' \href{http://dx.doi.org/10.1007/JHEP01(2013)142}{{\em
  JHEP} {\bf 1301} (2013)  142},
\href{http://arxiv.org/abs/1211.0019}{{\tt arXiv:1211.0019 [hep-th]}}.
%%CITATION = ARXIV:1211.0019;%%.

\bibitem{Hori:2000kt}
K.~Hori and C.~Vafa, ``{Mirror Symmetry},''
  \href{http://arxiv.org/abs/hep-th/0002222}{{\tt arXiv:hep-th/0002222}}.

\bibitem{Hori:2003ic}
K.~Hori, S.~Katz, A.~Klemm, R.~Pandharipande, R.~Thomas, C.~Vafa, R.~Vakil, and
  E.~Zaslow, {\em {Mirror Symmetry}}, vol.~1 of {\em {Clay Mathematics
  Monographs}}.
\newblock {Clay Mathematics Institute, Cambridge MA},
2003.
\newblock
%%CITATION = INSPIRE-640298;%%.

\bibitem{Crichigno:2015pma}
P.~M. Crichigno and M.~Roček, ``{On Gauged Linear Sigma Models with
  Torsion},'' \href{http://dx.doi.org/10.1007/JHEP09(2015)207}{{\em JHEP} {\bf
  1509} (2015)  207},
\href{http://arxiv.org/abs/1506.00335}{{\tt arXiv:1506.00335 [hep-th]}}.
%%CITATION = ARXIV:1506.00335;%%.

\bibitem{Merrell:2006py}
W.~Merrell, L.~A. Pando~Zayas, and D.~Vaman, ``{Gauged (2,2) Sigma Models and
  Generalized Kähler Geometry},''
  \href{http://dx.doi.org/10.1088/1126-6708/2007/12/039}{{\em JHEP} {\bf 0712}
  (2007)  039}, \href{http://arxiv.org/abs/hep-th/0610116}{{\tt
  arXiv:hep-th/0610116}}.

\bibitem{Lindstrom:2007vc}
U.~Lindström, M.~Roček, I.~Ryb, R.~von Unge, and M.~Zabzine, ``{New
  $\mathcal{N}{=} (2,2)$ Vector Multiplets},''
  \href{http://dx.doi.org/10.1088/1126-6708/2007/08/008}{{\em JHEP} {\bf 0708}
  (2007)  008},
\href{http://arxiv.org/abs/0705.3201}{{\tt arXiv:0705.3201 [hep-th]}}.
%%CITATION = ARXIV:0705.3201;%%.

\bibitem{Lindstrom:2007sq}
U.~Lindström, M.~Roček, I.~Ryb, R.~von Unge, and M.~Zabzine, ``{T-duality and
  Generalized Kähler Geometry},''
  \href{http://dx.doi.org/10.1088/1126-6708/2008/02/056}{{\em JHEP} {\bf 0802}
  (2008)  056},
\href{http://arxiv.org/abs/0707.1696}{{\tt arXiv:0707.1696 [hep-th]}}.
%%CITATION = ARXIV:0707.1696;%%.

\bibitem{Merrell:2007sr}
W.~Merrell and D.~Vaman, ``{T-duality, Quotients and Generalized Kähler
  Geometry},'' \href{http://dx.doi.org/10.1016/j.physletb.2008.06.031}{{\em
  Phys.Lett.} {\bf B665} (2008)  401--408},
\href{http://arxiv.org/abs/0707.1697}{{\tt arXiv:0707.1697 [hep-th]}}.
%%CITATION = ARXIV:0707.1697;%%.

\bibitem{Crichigno:2011aa}
P.~M. Crichigno, ``{The Semi-Chiral Quotient, Hyperkahler Manifolds and
  T-Duality},'' \href{http://dx.doi.org/10.1007/JHEP10(2012)046}{{\em JHEP}
  {\bf 1210} (2012)  046},
\href{http://arxiv.org/abs/1112.1952}{{\tt arXiv:1112.1952 [hep-th]}}.
%%CITATION = ARXIV:1112.1952;%%.

\bibitem{CrichignoThesis}
P.~M. Crichigno, {\em {Aspects of Supersymmetric Field Theories and Complex
  Geometry}}.
\newblock PhD thesis, {Stony Brook Univeristy}, 2013.

\bibitem{Kapustin:2006ic}
A.~Kapustin and A.~Tomasiello, ``{The General (2,2) Gauged Sigma Model with
  Three-form Flux},''
  \href{http://dx.doi.org/10.1088/1126-6708/2007/11/053}{{\em JHEP} {\bf 0711}
  (2007)  053}, \href{http://arxiv.org/abs/hep-th/0610210}{{\tt
  arXiv:hep-th/0610210}}.

\bibitem{Gates:1984nk}
S.~J. Gates~Jr., C.~M. Hull, and M.~Roček, ``{Twisted Multiplets and New
  Supersymmetric Nonlinear Sigma Models},''
\href{http://dx.doi.org/10.1016/0550-3213(84)90592-3}{{\em Nucl.Phys.} {\bf
  B248} (1984)  157}.
%%CITATION = NUPHA,B248,157;%%.

\bibitem{Gualtieri:2003dx}
M.~Gualtieri, {\em {Generalized Complex Geometry}}.
\newblock PhD thesis, Oxford University, 2003.
\newblock
\href{http://arxiv.org/abs/math/0401221}{{\tt arXiv:math/0401221 [math.DG]}}.
\newblock
%%CITATION = MATH/0401221;%%.

\bibitem{Gualtieri:2010fd}
M.~Gualtieri, ``{Generalized K\"ahler Geometry},''
  \href{http://dx.doi.org/10.1007/s00220-014-1926-z}{{\em Comm.Math.Phys.} {\bf
  331} (2014)  297--331}.

\bibitem{Zabzine:2006uz}
M.~Zabzine, ``{Lectures on Generalized Complex Geometry and Supersymmetry},''
  {\em Archivum Math.} {\bf 42} (2006)  119--146,
  \href{http://arxiv.org/abs/hep-th/0605148}{{\tt arXiv:hep-th/0605148}}.

\bibitem{Koerber:2010bx}
P.~Koerber, ``{Lectures on Generalized Complex Geometry for Physicists},''
  \href{http://dx.doi.org/10.1002/prop.201000083}{{\em Fortsch.Phys.} {\bf 59}
  (2011)  169--242},
\href{http://arxiv.org/abs/1006.1536}{{\tt arXiv:1006.1536 [hep-th]}}.
%%CITATION = ARXIV:1006.1536;%%.

\bibitem{Lindstrom:2005zr}
U.~Lindström, M.~Roček, R.~von Unge, and M.~Zabzine, ``{Generalized Kähler
  Manifolds and Off-shell Supersymmetry},''
  \href{http://dx.doi.org/10.1007/s00220-006-0149-3}{{\em Commun.Math.Phys.}
  {\bf 269} (2007)  833--849}, \href{http://arxiv.org/abs/hep-th/0512164}{{\tt
  arXiv:hep-th/0512164}}.

\bibitem{Ivanov:1994ec}
I.~T. Ivanov, B.~Kim, and M.~Roček, ``{Complex Structures, Duality and WZW
  Models in Extended Superspace},''
  \href{http://dx.doi.org/10.1016/0370-2693(94)01476-S}{{\em Phys.Lett.} {\bf
  B343} (1995)  133--143}, \href{http://arxiv.org/abs/hep-th/9406063}{{\tt
  arXiv:hep-th/9406063}}.

\bibitem{Lyakhovich:2002kc}
S.~Lyakhovich and M.~Zabzine, ``{Poisson Geometry of Sigma Models with Extended
  Supersymmetry},'' \href{http://dx.doi.org/10.1016/S0370-2693(02)02851-4}{{\em
  Phys.Lett.} {\bf B548} (2002)  243--251},
  \href{http://arxiv.org/abs/hep-th/0210043}{{\tt arXiv:hep-th/0210043}}.

\bibitem{Halmagyi:2007ft}
N.~Halmagyi and A.~Tomasiello, ``{Generalized Kähler Potentials from
  Supergravity},'' \href{http://dx.doi.org/10.1007/s00220-009-0881-6}{{\em
  Commun.Math.Phys.} {\bf 291} (2009)  1--30},
\href{http://arxiv.org/abs/0708.1032}{{\tt arXiv:0708.1032 [hep-th]}}.
%%CITATION = ARXIV:0708.1032;%%.

\bibitem{Kapustin:2004gv}
A.~Kapustin and Y.~Li, ``{Topological Sigma-models with $H$-flux and Twisted
  Generalized Complex Manifolds},'' {\em Adv.Theor.Math.Phys.} {\bf 11} (2007)
  269--290, \href{http://arxiv.org/abs/hep-th/0407249}{{\tt
  arXiv:hep-th/0407249}}.

\bibitem{Hitchin:2004ut}
N.~Hitchin, ``{Generalized Calabi-Yau Manifolds},''
  \href{http://dx.doi.org/10.1093/qjmath/54.3.281}{{\em Quart.J.Math.} {\bf 54}
  (2003)  281--308},
\href{http://arxiv.org/abs/math/0209099}{{\tt arXiv:math/0209099 [math.DG]}}.
%%CITATION = MATH/0209099;%%.

\bibitem{Grisaru:1997pg}
M.~T. Grisaru, M.~Massar, A.~Sevrin, and J.~Troost, ``{The Quantum Geometry of
  $\mathcal{N}{=}(2,2)$ Nonlinear Sigma Models},''
  \href{http://dx.doi.org/10.1016/S0370-2693(97)01053-8}{{\em Phys.Lett.} {\bf
  B412} (1997)  53--58}, \href{http://arxiv.org/abs/hep-th/9706218}{{\tt
  arXiv:hep-th/9706218}}.

\bibitem{Hull:2010sn}
C.~M. Hull, U.~Lindström, M.~Roček, R.~von Unge, and M.~Zabzine,
  ``{Generalized Calabi-Yau Metric and Generalized Monge-Ampere Equation},''
  \href{http://dx.doi.org/10.1007/JHEP08(2010)060}{{\em JHEP} {\bf 1008} (2010)
   060},
\href{http://arxiv.org/abs/1005.5658}{{\tt arXiv:1005.5658 [hep-th]}}.
%%CITATION = ARXIV:1005.5658;%%.

\bibitem{Hull:2008de}
C.~M. Hull, U.~Lindström, L.~Melo~dos Santos, R.~von Unge, and M.~Zabzine,
  ``{Euclidean Supersymmetry, Twisting and Topological Sigma Models},''
  \href{http://dx.doi.org/10.1088/1126-6708/2008/06/031}{{\em JHEP} {\bf 0806}
  (2008)  031},
\href{http://arxiv.org/abs/0805.3321}{{\tt arXiv:0805.3321 [hep-th]}}.
%%CITATION = ARXIV:0805.3321;%%.

\bibitem{Buscher:1987uw}
T.~Buscher, U.~Lindström, and M.~Roček, ``{New Supersymmetric $\sigma$ Models
  With {Wess-Zumino} Terms},''
\href{http://dx.doi.org/10.1016/0370-2693(88)90859-3}{{\em Phys.Lett.} {\bf
  B202} (1988)  94--98}.
%%CITATION = PHLTA,B202,94;%%.

\bibitem{Hori:2006dk}
K.~Hori and D.~Tong, ``{Aspects of Non-Abelian Gauge Dynamics in
  Two-Dimensional $\mathcal{N}{=}(2,2)$ Theories},''
  \href{http://dx.doi.org/10.1088/1126-6708/2007/05/079}{{\em JHEP} {\bf 0705}
  (2007)  079}, \href{http://arxiv.org/abs/hep-th/0609032}{{\tt
  arXiv:hep-th/0609032}}.

\bibitem{Lindstrom:2008hx}
U.~Lindström, M.~Roček, I.~Ryb, R.~von Unge, and M.~Zabzine, ``{Nonabelian
  Generalized Gauge Multiplets},''
  \href{http://dx.doi.org/10.1088/1126-6708/2009/02/020}{{\em JHEP} {\bf 0902}
  (2009)  020},
\href{http://arxiv.org/abs/0808.1535}{{\tt arXiv:0808.1535 [hep-th]}}.
%%CITATION = ARXIV:0808.1535;%%.

\bibitem{Closset:2014pda}
C.~Closset and S.~Cremonesi, ``{Comments on $\mathcal{N}{=}(2, 2)$
  Supersymmetry on Two-manifolds},''
  \href{http://dx.doi.org/10.1007/JHEP07(2014)075}{{\em JHEP} {\bf 1407} (2014)
   075},
\href{http://arxiv.org/abs/1404.2636}{{\tt arXiv:1404.2636 [hep-th]}}.
%%CITATION = ARXIV:1404.2636;%%.

\bibitem{Festuccia:2011ws}
G.~Festuccia and N.~Seiberg, ``{Rigid Supersymmetric Theories in Curved
  Superspace},'' \href{http://dx.doi.org/10.1007/JHEP06(2011)114}{{\em JHEP}
  {\bf 1106} (2011)  114},
\href{http://arxiv.org/abs/1105.0689}{{\tt arXiv:1105.0689 [hep-th]}}.
%%CITATION = ARXIV:1105.0689;%%.

\bibitem{Bonelli:2013mma}
G.~Bonelli, A.~Sciarappa, A.~Tanzini, and P.~Vasko, ``{Vortex Partition
  Functions, Wall Crossing and Equivariant Gromov-Witten Invariants},''
  \href{http://dx.doi.org/10.1007/s00220-014-2193-8}{{\em Commun.Math.Phys.}
  {\bf 333} (2015)  717--760},
\href{http://arxiv.org/abs/1307.5997}{{\tt arXiv:1307.5997 [hep-th]}}.
%%CITATION = ARXIV:1307.5997;%%.

\bibitem{Bredthauer:2006hf}
A.~Bredthauer, U.~Lindström, J.~Persson, and M.~Zabzine, ``{Generalized
  Kähler Geometry from Supersymmetric Sigma Models},''
  \href{http://dx.doi.org/10.1007/s11005-006-0099-x}{{\em Lett.Math.Phys.} {\bf
  77} (2006)  291--308}, \href{http://arxiv.org/abs/hep-th/0603130}{{\tt
  arXiv:hep-th/0603130}}.

\bibitem{Chuang:2006vt}
W.-Y. Chuang, ``{Topological Twisted Sigma Model with $H$-flux Revisited},''
  \href{http://dx.doi.org/10.1088/1751-8113/41/11/115402}{{\em J.Phys.} {\bf
  A41} (2008)  115402}, \href{http://arxiv.org/abs/hep-th/0608119}{{\tt
  arXiv:hep-th/0608119}}.

\bibitem{Zucchini:2006ii}
R.~Zucchini, ``{The bi-Hermitian Topological Sigma Model},''
  \href{http://dx.doi.org/10.1088/1126-6708/2006/12/039}{{\em JHEP} {\bf 0612}
  (2006)  039}, \href{http://arxiv.org/abs/hep-th/0608145}{{\tt
  arXiv:hep-th/0608145}}.

\bibitem{Hull:2008vs}
C.~M. Hull, U.~Lindström, L.~Melo~dos Santos, R.~von Unge, and M.~Zabzine,
  ``{Topological Sigma Models with $H$-Flux},''
  \href{http://dx.doi.org/10.1088/1126-6708/2008/09/057}{{\em JHEP} {\bf 0809}
  (2008)  057},
\href{http://arxiv.org/abs/0803.1995}{{\tt arXiv:0803.1995 [hep-th]}}.
%%CITATION = ARXIV:0803.1995;%%.

\bibitem{Candelas:1989js}
P.~Candelas and X.~C. de~la Ossa, ``{Comments on Conifolds},''
\href{http://dx.doi.org/10.1016/0550-3213(90)90577-Z}{{\em Nucl.Phys.} {\bf
  B342} (1990)  246--268}.
%%CITATION = NUPHA,B342,246;%%.

\bibitem{Klebanov:1998hh}
I.~R. Klebanov and E.~Witten, ``{Superconformal Field Theory on Three-branes at
  a Calabi-Yau Singularity},''
  \href{http://dx.doi.org/10.1016/S0550-3213(98)00654-3}{{\em Nucl.Phys.} {\bf
  B536} (1998)  199--218}, \href{http://arxiv.org/abs/hep-th/9807080}{{\tt
  arXiv:hep-th/9807080}}.

\bibitem{Candelas:1990rm}
P.~Candelas, X.~C. de~la Ossa, P.~S. Green, and L.~Parkes, ``{A Pair of
  Calabi-Yau Manifolds as an Exactly Soluble Superconformal Theory},''
\href{http://dx.doi.org/10.1016/0550-3213(91)90292-6}{{\em Nucl.Phys.} {\bf
  B359} (1991)  21--74}.
%%CITATION = NUPHA,B359,21;%%.

\bibitem{Gopakumar:1998ki}
R.~Gopakumar and C.~Vafa, ``{On the Gauge Theory / Geometry Correspondence},''
  {\em Adv.Theor.Math.Phys.} {\bf 3} (1999)  1415--1443,
  \href{http://arxiv.org/abs/hep-th/9811131}{{\tt arXiv:hep-th/9811131}}.

\bibitem{Ooguri:1999bv}
H.~Ooguri and C.~Vafa, ``{Knot Invariants and Topological Strings},''
  \href{http://dx.doi.org/10.1016/S0550-3213(00)00118-8}{{\em Nucl.Phys.} {\bf
  B577} (2000)  419--438}, \href{http://arxiv.org/abs/hep-th/9912123}{{\tt
  arXiv:hep-th/9912123}}.

\bibitem{Cavalcanti:2007}
G.~R. Cavalcanti and M.~Gualtieri, ``{A Surgery for Generalized Complex
  Structures on 4-manifolds},'' {\em J.Diff.Geom.} {\bf 76} (2007)  35--43,
  \href{http://arxiv.org/abs/math/0602333}{{\tt arXiv:math/0602333 [math.DG]}}.

\bibitem{Cavalcanti:2008ur}
G.~R. Cavalcanti and M.~Gualtieri, ``{Blow-up of Generalized Complex
  4-manifolds},'' \href{http://dx.doi.org/10.1112/jtopol/jtp031}{{\em J.Topol.}
  {\bf 2} (2009)  840--864},
\href{http://arxiv.org/abs/0806.0872}{{\tt arXiv:0806.0872 [math.SG]}}.
%%CITATION = ARXIV:0806.0872;%%.

\bibitem{Grisaru:1997ep}
M.~T. Grisaru, M.~Massar, A.~Sevrin, and J.~Troost, ``{Some Aspects of
  $\mathcal{N}{=}(2,2)$, $D = 2$ Supersymmetry},'' {\em Fortsch.Phys.} {\bf 47}
  (1999)  301--307, \href{http://arxiv.org/abs/hep-th/9801080}{{\tt
  arXiv:hep-th/9801080}}.

\bibitem{Bogaerts:1999jc}
J.~Bogaerts, A.~Sevrin, S.~van~der Loo, and S.~Van~Gils, ``{Properties of
  Semichiral Superfields},''
  \href{http://dx.doi.org/10.1016/S0550-3213(99)00490-3}{{\em Nucl.Phys.} {\bf
  B562} (1999)  277--290}, \href{http://arxiv.org/abs/hep-th/9905141}{{\tt
  arXiv:hep-th/9905141}}.

\bibitem{Gualtieri:2007ng}
M.~Gualtieri, ``{Generalized Complex Geometry},''
  \href{http://dx.doi.org/10.4007/annals.2011.174.1.3}{{\em {Ann.Math.}} {\bf
  174} (2011)  75--123}, \href{http://arxiv.org/abs/math/0703298}{{\tt
  arXiv:math/0703298 [math.DG]}}.

\bibitem{Sevrin:2011mc}
A.~Sevrin, W.~Staessens, and D.~Terryn, ``{The Generalized Kähler Geometry of
  $\mathcal{N}{=}(2,2)$ WZW-models},''
  \href{http://dx.doi.org/10.1007/JHEP12(2011)079}{{\em JHEP} {\bf 1112} (2011)
   079},
\href{http://arxiv.org/abs/1111.0551}{{\tt arXiv:1111.0551 [hep-th]}}.
%%CITATION = ARXIV:1111.0551;%%.

\end{thebibliography}\endgroup

\end{document}